\documentclass[nofootinbib,floatfix,onecolumn,superscriptaddress,a4paper]{revtex4}

\pdfoutput=1

\usepackage[colorlinks=true,citecolor=blue,linkcolor=purple,urlcolor=purple]{hyperref}

\usepackage{amsmath}
\usepackage{amssymb}
\usepackage{graphicx}
\usepackage{xcolor}
\usepackage[T1]{fontenc} 
\usepackage{slashed}
\usepackage{enumitem}

\renewcommand{\arraystretch}{1.6}
\newcommand{\hc}{\mathrm{h.c.}}

\newcommand{\id}{\mathbb{I}}
\newcommand{\npar}{\texttt{\#}}

\def\gsim{\raise0.3ex\hbox{$\;>$\kern-0.75em\raise-1.1ex\hbox{$\sim\;$}}}
\def\lsim{\raise0.3ex\hbox{$\;<$\kern-0.75em\raise-1.1ex\hbox{$\sim\;$}}}

\newcommand{\AddrIFIC}{%
Instituto de F\'{\i}sica Corpuscular (CSIC-Universitat de Val\`{e}ncia),
Apdo. 22085, E-46071 Valencia, Spain
}

\newcommand{\AddrUV}{%
Departamento de Matem\'aticas, Universitat de Val\`encia, E-46100 Burjassot, 
Val\`encia, Spain}

\preprint{IFIC/19-59}
\begin{document}

\title{General parametrization of Majorana neutrino mass models \\
\vspace*{0.2cm}
{\normalsize One formula to fit them all!}}

\author{I. Cordero-Carri\'on}\email{isabel.cordero@uv.es}
\affiliation{\AddrUV}

\author{M. Hirsch} \email{mahirsch@ific.uv.es }
\affiliation{\AddrIFIC}

\author{A. Vicente} \email{avelino.vicente@ific.uv.es}
\affiliation{\AddrIFIC}

\begin{abstract}
We discuss a general formula which allows to automatically reproduce
experimental data for Majorana neutrino mass models, while keeping the
complete set of the remaining model parameters free for general scans,
as necessary in order to provide reliable predictions for observables
outside the neutrino sector.  We provide a proof of this master
parametrization and show how to apply it for several well-known
neutrino mass models from the literature. We also discuss a list of
special cases, in which the Yukawa couplings have to fulfill some
particular additional conditions.

\end{abstract}

\maketitle

\section{Introduction}
\label{sec:intro}

Most of the classical Majorana neutrino mass models, such as the three
tree-level seesaws
(type-I~\cite{Minkowski:1977sc,Yanagida:1979as,Mohapatra:1979ia,GellMann:1980vs,Schechter:1980gr},
type-II~\cite{Mohapatra:1980yp,Schechter:1980gr} and
type-III~\cite{Foot:1988aq}) or the 1-loop Zee model \cite{Zee:1980ai}
and the 2-loop Babu-Zee
model~\cite{Cheng:1980qt,Zee:1985id,Babu:1988ki} have all been
discussed already in the 1980's. However, ever since the discovery of
neutrino oscillations~\cite{Fukuda:1998mi,Ahmad:2002jz} a miryad more
of other neutrino mass models has been proposed in the literature.

To name a few papers and reviews post-1998, we mention
\cite{Ma:1998dn}, which showed that there are only three types of
seesaws at tree-level.  For a systematic analysis of all possible
1-loop diagrams, see \cite{Bonnet:2012kz}. At 2-loop level we mention
two different \textit{colored} versions of the Babu-Zee topology
\cite{Babu:2011vb,Angel:2013hla}. A general decomposition for all
2-loop models was presented in \cite{Sierra:2014rxa}. At three-loop
order there are the KNT~\cite{Krauss:2002px}, AKS~\cite{Aoki:2008av}
and cocktail models~\cite{Gustafsson:2012vj}.  And, recently, for
3-loops a systematic analysis was given in \cite{Cepedello:2018rfh}.
One can find even some examples of 4-loop models in the literature
\cite{Gu:2011ak,Helo:2015fba}.  For a recent review on radiative
neutrino mass models, we refer to \cite{Cai:2017jrq}.

One of the basic problems faced by model builders is to first
reproduce correctly the measured neutrino masses and angles and to
then scan over all remaining free parameters of the model in a
systematic way, in order to explore possible predictions the model may
make for other observables, such as $\mu\to e\gamma$ or neutrinoless
double beta decay.  It is often not difficult to identify some
singular point in the parameter space of a given model, which explains
oscillation data. However, exploring the parameter space in a complete
and un-biased way seems not to be straight-foward in many cases.
Here, we discuss in detail the master formula for neutrino mass
models, introduced first in~\cite{Cordero-Carrion:2018xre}. All
Majorana neutrino mass models can be brought to this form. We then
discuss the master parametrization, a specific set of equations which
allow to solve the above problem in a systematic way.

This paper is organized as follows. In Section~\ref{sec:master} we
discuss the master parametrization. We define all necessary matrices
for the different possible cases and show by explicit parameter
counting that the complete parameter space of any given model can be
covered in this way.  We then turn to a discussion of how to apply our
general master parametrization for some specific example models. We
start with the simplest type-I seesaw
model~\cite{Minkowski:1977sc,Yanagida:1979as,Mohapatra:1979ia,GellMann:1980vs,Schechter:1980gr}
and demonstrate how our general parametrization can be reduced to the
well-known Casas-Ibarra parametrization~\cite{Casas:2001sr} for this
case. In increasing order of complexity, we then discuss the inverse
seesaw~\cite{Mohapatra:1986bd}, the scotogenic model~\cite{Ma:2006km}
(as an example of a radiative model) and finally the linear
seesaw~\cite{Akhmedov:1995ip,Akhmedov:1995vm}.

We then turn to discuss a list of special cases. These are models in
which some Yukawa matrices are not completely free parameters, but for
theoretical reasons have to fulfill some particular conditions, such as
$y\equiv y^T$, as happens, for example, in left-right symmetric models.
Constraints on Yukawa matrices appear in many more models, in
particular models with family symmetries are of this type. For a review
on neutrino mass models with discrete symmetries, see for example
\cite{King:2013eh}.  We demonstrate, how our general formalism can be
adapted to such additional conditions for several cases and we also
discuss the limitations of our approach: while our master
parametrization is valid for all cases, solving the equations may
become impractically complicated, if there are too many additional
conditions.

We then close with a short summary. A number of more technical aspects
of our work is discussed in appendices. Appendix~\ref{sec:proof} gives
the proof of our master parametrization. Appendix~\ref{sec:mat}
provides specific parametrizations for some of the matrices involved
in the master parametrization. Appendices~\ref{sec:proof_antisym1} and
\ref{sec:proof_antisym2} discuss the master parametrization in the
special cases with one or two antisymmetric Yukawa
matrices. Appendix~\ref{sec:finetuning} demonstrates in one concrete
example model, how to account for higher-order corrections in
particular corners of parameter space, where the parameters in the
leading order contribution are particularly fine-tuned. Finally, in
Appendix~\ref{sec:hybrid} we discuss how to apply our general equation
to scenarios with several contributions to the neutrino mass matrix.

\section{The master formula and parametrization}
\label{sec:master}

\subsection{General neutrino mass matrix}
\label{subsec:massmat}

The contributions from any Majorana neutrino
mass model can be brought into the form:
\begin{equation} \label{eq:master}
m = f \, \left( y_1^T \, M \, y_2 + y_2^T \, M^T \, y_1 \right) \, .
\end{equation}
$m$ is a complex symmetric matrix. Since there are 3 generations of
light, active neutrinos we assume it has dimensions $3 \times 3$, but
it is straight-forward to generalize all equations below to more 
generations. $m$ can be brought to diagonal form 
using a Takagi decomposition as
\begin{equation} \label{eq:m-U}
D_m = \text{diag} \left( m_1, m_2, m_3 \right) = U^T \, m \, U \, ,
\end{equation}
where $U$ is a $3 \times 3$ unitary matrix ($U^\dagger U = U U^\dagger
= \id_3$).~\footnote{The matrices $D_m$ and $U$ are strongly connected
  to neutrino oscillation experiments, as explained in
  Appendix~\ref{sec:mat}. We will assume $U$ to be a unitary matrix,
  thus neglecting possible non-unitarity effects, which are
  nevertheless experimentally constrained to be small.} The matrices
$y_1$ and $y_2$ in Eq.~\eqref{eq:master} are dimensionless $n_1 \times
3$ and $n_2 \times 3$ complex matrices, in general without any
symmetry restrictions.  $M$ is a $n_1 \times n_2$ complex matrix, with
dimension of mass.  In the following we assume without loss of
generality $n_1\geq n_2$.  Neutrino oscillation data requires that $m$
must contain at least two non-vanishing eigenvalues.  Therefore, we
concentrate on the cases $r_m=\text{rank}(m)=2\, \text{or}\, 3$.  We
treat both neutrino mass orderings: Normal Hierarchy (NH) and Inverted
Hierarchy (IH).

\subsection{Master parametrization}
\label{subsec:param}

We call Eq.~\eqref{eq:master} the \textit{master formula}, since it is
valid for all Majorana neutrino mass models. We now proceed to
discuss a parametrization for the $y_1$ and $y_2$ Yukawa matrices with
three specific properties:
\begin{itemize}
\item {\bf General:} valid for all models.
\item {\bf Complete:} containing all the degrees of freedom in the model.
\item {\bf Programmable:} easy to use in phenomenological analyses.
\end{itemize}

This parametrization of the Yukawa matrices will be called the
\textit{master parametrization}. As shown in Appendix~\ref{sec:proof}, 
the Yukawa matrices $y_1$ and $y_2$ can be parametrized in general 
as
\begin{align}
y_1 & = \frac{1}{\sqrt{2 \, f}} \, V_1^\dagger \, \left( \begin{array}{c}
\Sigma^{-1/2} \, W \, A \\ X_1 \\ X_2 \end{array} \right) \, 
\bar{D}_{\sqrt{m}} \, U^\dagger \, , \label{eq:par1} \\
y_2 & = \frac{1}{\sqrt{2 \, f}} \, V_2^\dagger \, \left( \begin{array}{c}
\Sigma^{-1/2} \, \widehat W^\ast \, \widehat B \\ X_3 \end{array} \right) 
\, \bar{D}_{\sqrt{m}} \, U^\dagger \, . \label{eq:par2}
\end{align}
Here, $\ast$ denotes complex conjugation and $\dagger$ hermitian 
conjugation as usual. The matrix $\bar{D}_{\sqrt{m}}$ is defined as
\begin{equation} \label{eq:defDsqrtm}
\bar{D}_{\sqrt{m}} = \left\{ \begin{array}{ll}
\text{diag}\left(\sqrt{m_1},\sqrt{m_2},\sqrt{m_3}\right) \quad & \text{if} \ r_m=3 \, , \\
P \cdot \text{diag}\left(\sqrt{v},\sqrt{m_2},\sqrt{m_3}\right) \cdot P \quad & \text{if} \ r_m=2 \, .
\end{array} \right.
\end{equation}
with
\begin{equation} \label{eq:Pdef}
P = \left\{ \begin{array}{cl}
\id_3 \quad & \text{for NH} \, , \\
P_{13} \quad & \text{for IH} \, ,
\end{array} \right.
\end{equation}
and
\begin{equation} \label{eq:P13}
P_{13} = \begin{pmatrix} 0 & 0 & 1 \\ 0 & 1 & 0 \\ 1 & 0 & 0 \end{pmatrix} \, ,
\end{equation}
a permutation matrix. We note that our definition of
$\bar{D}_{\sqrt{m}}$ in case of $r_m = 2$ adopts the standard form in
case of NH by choosing $P = \id_3$. The form
$\text{diag}\left(\sqrt{m_1},\sqrt{m_2},\sqrt{v}\right)$, more
commonly used in case of IH, is obtained by choosing $P = P_{13}$ and
then renaming $m_3 \to m_1$. The scale $v$ can be replaced in this
definition by any non-vanishing reference mass scale.~\footnote{It may
  naively seem that the $\sqrt{v}$ entry in the definition of
  $\bar{D}_{\sqrt{m}}$ in Eq.~\eqref{eq:defDsqrtm} is a free
  parameter. However, this is not the case. Even though this entry
  will appear explicitly in the analytical expressions of $y_1$ and
  $y_2$ when $r_m = 2$, it is easy to see that a change in this
  parameter can be absorbed by rescaling the first (third) column of
  $T$ and the first (third) row of $K$, two matrices to be defined
  below, in case of NH (IH). Therefore, the freedom in this entry is
  already \textit{covered} by the $T$ and $K$ matrices, when their
  elements are considered in their complete domains. In summary, this
  entry does not add any free parameter to the master parametrization
  and one can fix it to a specific value. We chose $\sqrt{v}$, with
  $v$ the usual electroweak vacuum expectation value, for
  simplicity. Finally, we note that this scale, although arbitrary,
  cannot vanish. This would imply $T$ and $K$ matrices out of their
  ranges of validity, a fact that is reflected in the proof given in
  Appendix~\ref{sec:proof}, where the existence of
  $\bar{D}_{\sqrt{m}}^{-1}$ is required.} We applied a singular-value
decomposition to the matrix $M$,
\begin{equation} \label{eq:M-SVD}
M = V_1^T \, \widehat \Sigma \, V_2 \, ,
\end{equation}
where $\widehat \Sigma$ is a $n_1 \times n_2$ matrix defined as 
\begin{equation} \label{eq:sigma}
\widehat \Sigma = \left(\begin{array}{c} \begin{array}{cc} 
\Sigma & 0 \\ 0 & 0_{n_2-n} \end{array} \\ \hline 0_{n_1-n_2} 
\end{array}\right) \, ,
\end{equation}
and $\Sigma =\text{diag}\left(\sigma_1,\sigma_2,\dots,\sigma_n\right)$
is a diagonal $n \times n$ matrix containing the positive and real
singular values of $M$ ($\sigma_i>0$). $M$ can have vanishing singular
values which we encode in the zero square $(n_2-n)\times(n_2-n)$
matrix $0_{n_2-n}$. $V_1$ and $V_2$ are $n_1 \times n_1$ and $n_2
\times n_2$ unitary matrices, which can be found by diagonalizing the
square matrices $M M^\dagger$ and $M^\dagger M$, respectively. $X_1$,
$X_2$ and $X_3$ are, respectively, $(n_2-n)\times 3$, $(n_1-n_2)\times
3$ and $(n_2-n)\times 3$ arbitrary complex matrices with dimensions of
mass$^{-1/2}$. $\widehat W$ is an $n\times n$ matrix defined as
\begin{equation}
\widehat W = \left(W \quad \bar{W}\right) \, ,
\end{equation}
where $W$ is an $n\times r$ complex matrix, with $r=\text{rank}(W)$,
such that $W^\dagger W = W^T W^* = \id_r$, while $\bar{W}$ is an
$n\times(n-r)$ complex matrix, that is built with vectors that complete 
those in $W$ to form an orthonormal basis of $\mathbb{C}^n$. Thus, 
$\widehat W$ is a complex unitary $n\times n$ matrix. A specific form for this matrix can be found in Appendix~\ref{sec:mat}. $A$ is given 
as a $r\times 3$ matrix, which can in general be written as
\begin{equation}
A = T \, C_1 \, ,
\end{equation}
where $T$ is an upper-triangular $r\times r$ invertible square
matrix with positive real values in the diagonal, and $C_1$ is an
$r\times 3$ matrix. Finally, $\widehat B$ is defined as a $n\times 3$
complex matrix given by
\begin{equation}
\widehat B = \left( \begin{array}{c} B \\ \bar{B} \\ \end{array} \right) \, ,
\end{equation}
with $\bar{B}$ an arbitrary $(n-r)\times 3$ complex matrix and $B$ an
$r\times 3$ complex matrix written as:
\begin{equation} \label{eq:Bexp}
B \equiv B\left( T , K , C_1 , C_2 \right) 
= \left( T^T \right)^{-1} \, \left[ C_1 \, C_2 + K \, C_1 \right] \, ,
\end{equation}
where we have introduced the antisymmetric $r\times r$ square matrix
$K$ and the $3\times 3$ matrix $C_2$.~\footnote{Eq.~\eqref{eq:master}
  shows that it is possible to scale up one of the two Yukawa matrices
  by a global factor $F$ and compensate it by inverse scaling of the
  other Yukawa by $1/F$. This freedom is of course taken into account
  in the master parametrization of Eqs. \eqref{eq:par1} and
  \eqref{eq:par2}. Multiplying $y_1$ by adding a factor in the matrix
  $T$, which enters $y_1$ via $A$, this factor will be exactly
  canceled out by that coming from $\left( T^T \right)^{-1}$ in $B$,
  see Eq.~\eqref{eq:Bexp}.} In the following $i = \sqrt{-1}$ is the
imaginary unit, as usual. The form of the matrices $C_1$ and $C_2$ is
case-dependent.  For different values of $r_m$ and $r$ they are given
as follows:~\footnote{The expression for $C_2$ in the $(3,3)$ case has
  been simplified with respect to \cite{Cordero-Carrion:2018xre}.}

\begin{itemize}
\item {\bf Case $\boldsymbol{(3,3)}$: $\boldsymbol{r_m = 3}$ and $\boldsymbol{r = 3}$:}
\end{itemize}

\begin{align}
C_1 = C_2 = \id_3 \, .
\end{align}

\begin{itemize}
\item {\bf Case $\boldsymbol{(3,2)}$: $\boldsymbol{r_m = 3}$ and $\boldsymbol{r = 2}$:}
\end{itemize}

In this case we find two sub-cases: case $(3,2)_a$, when the second
and third columns of the product matrix $W \, A$ are linearly
independent, and $(3,2)_b$, when they are linearly dependent. The
matrices $C_1$ and $C_2$ take the following expressions:

\begin{itemize}
\item Case $(3,2)_a$:
\end{itemize}

\begin{align} \label{eq:C1C232a}
C_1 = \left( \begin{array}{ccc}
z_1 & 1 & 0 \\
z_2 & 0 & 1 \end{array} \right) \, , \quad \text{with} \, 1 + z_1^2 + z_2^2 = 0, \quad
C_2 = \left( \begin{array}{ccc}
-1 & 0 & 0 \\
0 & 1 & 0 \\
0 & 0 & 1 \end{array} \right).
\end{align}
Here, $z_1$ and $z_2$ are complex numbers.

\begin{itemize}
\item Case $(3,2)_b$:
\end{itemize}

\begin{align} \label{eq:C1C232b}
C_1 = C_{1 \, \pm} = \left( \begin{array}{ccc}
0 & \pm i & 1 \\
1 & 0 & 0 \end{array} \right) 
\, , \quad
C_2 = \left( \begin{array}{ccc}
1 & 0 & 0 \\
0 & -1 & 0 \\
0 & 0 & 1 \end{array} \right) 
\, .
\end{align}

\begin{itemize}
\item {\bf Case $\boldsymbol{(2,3)}$: $\boldsymbol{r_m = 2}$ and $\boldsymbol{r = 3}$:}
\end{itemize}

\begin{align}
C_1 = P \, ,  \quad
C_2 = P \, \left( \begin{array}{ccc}
0 & 0 & 0 \\
0 & 1 & 0 \\
0 & 0 & 1 \end{array} \right) \, P \, .
\end{align}

\begin{itemize}
\item {\bf Case $\boldsymbol{(2,2)}$: $\boldsymbol{r_m = 2}$ and $\boldsymbol{r = 2}$:}
\end{itemize}

In this case we again sub-divide into two sub-cases: case $(2,2)_a$,
when the second and third columns of the matrix $W \, A$ are linearly
independent, and $(2,2)_b$, when they are linearly dependent. The
matrices $C_1$ and $C_2$ take the following expressions:

\begin{itemize}
\item Case $(2,2)_a$:
\end{itemize}

\begin{align} \label{eq:C1C222a}
C_1 = \left( \begin{array}{ccc}
z_1 & 1 & 0 \\
z_2 & 0 & 1 \end{array} \right) \, P \, , \quad \text{with} \, z_1^2 + z_2^2 = 0, \quad
C_2 = P \, \left( \begin{array}{ccc}
-1 & 0 & 0 \\
0 & 1 & 0 \\
0 & 0 & 1 \end{array} \right) \, P \, .
\end{align}

\begin{itemize}
\item Case $(2,2)_b$:
\end{itemize}

\begin{align} \label{eq:C1C222b}
C_1 = C_{1 \, \pm} = \left( \begin{array}{ccc}
0 & \pm i & 1 \\
1 & 0 & 0 \end{array} \right) \, P
\, , \quad
C_2 = P \, \left( \begin{array}{ccc}
0 & 0 & 0 \\
0 & -1 & 0 \\
0 & 0 & 1 \end{array} \right) \, P
\, .
\end{align}

\begin{itemize}
\item {\bf Case $\boldsymbol{(2,1)}$: $\boldsymbol{r_m = 2}$ and $\boldsymbol{r = 1}$:}
\end{itemize}

We would like to point out that one can have two non-vanishing
eigenvalues in $m$ even for $r=1$ due to the fact that
Eq. \eqref{eq:master} has two terms contributing. In this case we note
that $K = 0_{1 \times 1}$. The matrices $C_1$ and $C_2$ take the
following expressions:

\begin{align}
C_1 = C_{1 \, \pm} = \left( \begin{array}{ccc}
0 & \pm i & 1 \end{array} \right) \, P \, , \quad
C_2 = P \, \left( \begin{array}{ccc}
0 & 0 & 0 \\
0 & -1 & 0 \\
0 & 0 & 1 \end{array} \right) \, P \, .
\end{align}

It can be shown that $(r_m,r)$ cases not considered here cannot be
made compatible with neutrino oscillation data and the master Majorana
mass matrix in Eq.~\eqref{eq:master}. We give a summary of the
matrices that appear in the master parametrization and count their
free parameters in Tab.~\ref{tab:matrices}. A rigorous mathematical
proof of the master parametrization is given in
Appendix~\ref{sec:proof}. Finally, a {\tt Mathematica} notebook that
implements the master parametrization can be found
in~\cite{masterweb}.

{
\renewcommand{\arraystretch}{1.4}
\begin{table*}[t]
\centering
{\setlength{\tabcolsep}{2em}
\begin{tabular}{| c c c c |}
\hline  
Matrix & Dimensions & Property & Real parameters \\
\hline
\hline    
$X_1$ & $(n_2-n) \times 3$ & Absent if $n = n_2$ & $6 \, (n_2-n)$ \\
$X_2$ & $(n_1-n_2) \times 3$ & Absent if $n_1 = n_2$ & $6 \, (n_1-n_2)$ \\
$X_3$ & $(n_2-n) \times 3$ & Absent if $n = n_2$ & $6 \, (n_2-n)$ \\
$W$ & $n \times r$ &  & $r \, (2n-r)$ \\
$T$ & $r \times r$ & Upper triangular with $(T)_{ii} > 0$ & $r^2$ \\
$K$ & $r \times r$ & Antisymmetric & $r \, (r-1)$ \\
$\bar B$ & $(n-r) \times 3$ & Absent if $n = r$ & $6 \, (n-r)$ \\
$C_1$ & $r \times 3$ & Case-dependent & $0$ or $2$ \\
$C_2$ & $3 \times 3$ & Case-dependent & - \\
\hline
\end{tabular}
}
\caption{Matrices containing free parameters in the master
  parametrization. Even though the matrix $C_2$ does not contain any
  free parameter, we include it in this list since its form depends on
  the values of $r_m$ and $r$.
\label{tab:matrices}
}
\end{table*}
}

\subsection{Parameter counting}
\label{sec:counting}

Without loss of generality we can write
\begin{equation} \label{eq:counting}
\npar_{\rm free} = \npar_{y_1} + \npar_{y_2} - \npar_{\rm eqs} - \npar_{\rm extra}
= 6 (n_1 + n_2) - \npar_{\rm eqs} - \npar_{\rm extra} \, .
\end{equation}
Here $\npar_{y_1} = 2 \cdot 3 \cdot n_1$ and $\npar_{y_2} = 2 \cdot 3
\cdot n_2$ are the number of {\em real} degrees of freedom in $y_1$ and
$y_2$.  $\npar_{\rm eqs}$ is the number of
real independent equations contained in Eq. \eqref{eq:master}. Because 
this matrix equation is symmetric, the naive expectation is to have 6
complex equations. This would then correspond to 12 real
restrictions on the elements of $y_1$ and $y_2$. However, by direct 
computation one can show that for $r=1$ one of the complex
equations is redundant and can be derived from the other
five. Thus, 
\begin{equation}
\npar_{\rm eqs} = \left\{ \begin{array}{l}
12 \quad \text{for} \, r=3 \, \text{or} \, 2 ,\\
10 \quad \text{for} \, r=1 .
\end{array} \right.
\end{equation}
The case $r=1$ is actually allowed only because \eqref{eq:master} 
contains two terms. Each of these, in principle, can be of 
rank 1, as long as the rank of the sum of both terms is 2.
Finally, $\npar_{\rm extra}$ counts the number of extra (real)
restrictions imposed on $y_1$ and $y_2$. Often, such as in the case of
the minimal type-I seesaw, one has $\npar_{\rm extra} = 0$. However,
there are also many scenarios with additional restrictions and
$\npar_{\rm extra} \neq 0$. Since the number of free parameters
$\npar_{\rm free}$ must equal the sum of the number of free parameters
in each of the matrices, contained in the master parametrization of
Eqs. \eqref{eq:par1} and \eqref{eq:par2}, we find
\begin{align} 
\npar_{\rm free} &= \npar_{X_1} + \npar_{X_2} 
+ \npar_{X_3} + \npar_{A} + \npar_{W} 
+ \npar_{B} + \npar_{\bar{B}} + \npar_{C_1} 
\nonumber \\
&= \npar_{X_1} + \npar_{X_2} 
+ \npar_{X_3} + \npar_{T} + \npar_{W} 
+ \npar_{K} + \npar_{\bar{B}} + \npar_{C_1} \, . \label{eq:counting2}
\end{align}
In these expressions we assigned all the free parameters in the
product $\bar{W} \, \bar{B}$ to $\bar{B}$, corresponding to
$\npar_{\bar{W}} = 0$.  We can always choose this, since these two
matrices appear everywhere in the combination $\bar{W} \,
\bar{B}$. Considering that all the parameters contained in $\bar{B}$
are free, $\npar_{\bar{W}\bar{B}} \equiv \npar_{\bar{B}}$. Next, one can easily count the parameters in
each of the matrices in Eq.~\eqref{eq:counting2} and find
\begin{align}
\npar_{X_1} &= 6 \, (n_2-n) \, , \nonumber \\
\npar_{X_2} &= 6 \, (n_1-n_2) \, , \nonumber \\
\npar_{X_3} &= 6 \, (n_2-n) \, , \nonumber \\
\npar_T &= r^2 \, , \nonumber \\
\npar_K &= r \, (r-1) \, , \nonumber \\
\npar_{\bar{B}} &= 6 \, (n-r) \, . \label{eq:allcounting}
\end{align}
The counting of the free parameters in $W$ is more involved, but it
can be found by constructing a set of $r$ orthonormal vectors of $n$
components, and counting the number of conditions that orthonormality
imposes on them. One finds
\begin{equation} \label{eq:Wcounting}
\npar_W = r \, (2n-r) \, .
\end{equation}
Finally, we note that $\npar_{C_1} = 0$ in most cases, except for
cases $(3,2)_a$ and $(2,2)_a$, for which $\npar_{C_1}=2$.  The
parameter counting for the matrices in the master parametrization is
shown in Table~\ref{tab:matrices}. For pedagogical and practical
purposes, we also provide Table~\ref{tab:counting}, where we detail
the number of free parameters for several selected scenarios and how
they distribute among the different matrices.

\begin{table}
\centering
\begin{tabular}{|c|cccccc|c|cccccccc|}
\hline
Scenario & $n_1$ & $n_2$ & $n$ & case & $\npar_{\rm eqs}$ & $\npar_{\rm extra}$ & $\npar_{\rm free}$ & $\npar_{X_1}$ & $\npar_{X_2}$ & $\npar_{X_3}$ & $\npar_{T}$ & $\npar_{W}$ & $\npar_{K}$ & $\npar_{\bar{B}}$ & $\npar_{C_1}$ \\
\hline
1 & 3 & 3 & 3 & $(3,3)$ & 12 & 0 & 24 & - & - & - & 9 & 9 & 6 & - & - \\
2 & 4 & 3 & 2 & $(3,3)$ & 12 & 0 & 42 & 6 & 6 & 6 & 9 & 9 & 6 & - & - \\
3 & 3 & 3 & 3 & $(3,2)_a$ & 12 & 2 & 22 & - & - & - & 4 & 8 & 2 & 6 & 2 \\
4 & 2 & 2 & 2 & $(3,2)_a$ & 12 & 0 & 12 & - & - & - & 4 & 4 & 2 & - & 2 \\
5 & 3 & 3 & 3 & $(3,2)_b$ & 12 & 4 & 20 & - & - & - & 4 & 8 & 2 & 6 & - \\
6 & 2 & 2 & 2 & $(2,2)_a$ & 12 & 0 & 12 & - & - & - & 4 & 4 & 2 & - & 2 \\
7 & 2 & 2 & 2 & $(2,2)_b$ & 12 & 2 & 10 & - & - & - & 4 & 4 & 2 & - & - \\
8 & 2 & 2 & 2 & $(2,1)$ & 10 & 4 & 10 & - & - & - & 1 & 3 & - & 6 & - \\
\hline
\end{tabular}
\caption{Parameter counting table. Here we detail the number of free
  parameters for some selected scenarios and how they distribute among
  the different matrices appearing in the master parametrization.
\label{tab:counting}}
\end{table}

It may be convenient to discuss the following particular example in
order to understand the general parameter counting procedure. Let us
choose $n_1 = n_2 = n = 3$ and consider on a scenario with $(r_m,r) =
(3,3)$.  Then, $\widehat \Sigma \equiv \Sigma$, $\npar_{\rm eqs} = 12$
and $\npar_{\rm extra} = 0$. From Eq.\eqref{eq:counting}, one
calculates $\npar_{\rm free}^{(3,3)}=24$. Applying now
Eq. \eqref{eq:counting2}, one finds
\begin{equation}
\npar_{\rm free}^{(3,3)} = 24 = \npar_W^{(3,3)} + \npar_A^{(3,3)} 
+ \npar_B^{(3,3)} + \npar_{C_1}^{(3,3)} 
= 15 + \npar_{W}^{(3,3)} \, ,
\end{equation}
where $\npar_{W}^{(3,3)} = 9$ corresponds to the number of real free
parameters in the matrix $W$ in the $(3,3)$ case. We note that
$\npar_{W}^{(3,3)} = 9$ also follows from the fact that $W$ is a
unitary $3 \times 3$ matrix. This provides a consistency check of
the parameter counting we just demonstrated. In addition, note also 
$\npar_A^{(3,3)} = 9$ and $\npar_B^{(3,3)} = 6$.

\section{Example applications}
\label{sec:apps}

The practical use of the master parametrization is straightforward. It
can be easily applied to any Majorana neutrino mass model and
completely automatized in order to run detailed numerical
analyses. First, one must use the information from neutrino
oscillation experiments, typically from a global fit, and fix the
light neutrino masses and leptonic mixing angles appearing in $\bar
D_{\sqrt{m}}$ and $U$, respectively. In a second step one must compare
the expression for the mass matrix of the light neutrinos in the model
under consideration with the general master formula in
Eq.~\eqref{eq:master}. This way one can easily identify the global
factor $f$, the Yukawa matrices $y_1$ and $y_2$ as well as the matrix
$M$. The latter can be singular-value decomposed to determine
$\Sigma$, $V_1$ and $V_2$, while the Yukawa matrices $y_1$ and $y_2$
can be expressed in terms of a set of matrices ($\widehat W$,
$X_{1,2,3}$, $\bar B$, $T$, $K$ and $C_{1,2}$) by means of the master
parametrization in Eqs.~\eqref{eq:par1} and \eqref{eq:par2}. Finally,
in a numerical analysis one can simply randomly scan over the free
parameters contained in these matrices to completely explore the
parameter space of a given model. \\

We will now illustrate the use of the master parametrization with
several example models. In the following, $H$ will denote the SM Higgs
doublet, transforming as $({\bf 1}, {\bf 2}, 1/2)$ under the SM gauge
symmetry, whereas $L$ will denote the SM lepton doublets, transforming
as $({\bf 1}, {\bf 2}, -1/2)$, and $e_R$ the SM lepton singlets,
transforming as $({\bf 1}, {\bf 1}, -1)$. As already mentioned in 
section \ref{sec:master} we will work in the basis, where the 
charged lepton mass matrix has already been diagonalized. 

\subsection{The type-I seesaw}
\label{subsec:typeI}

{
\renewcommand{\arraystretch}{1.4}
\begin{table}
\centering
{\setlength{\tabcolsep}{0.5em}
\begin{tabular}{| c c c c c c |}
\hline  
 & spin & generations & $\mathrm{SU(3)}_c$ & $\mathrm{SU(2)}_L$ & $\mathrm{U(1)}_Y$ \\
\hline
\hline    
$N$ & 1/2 & 3 & ${\bf 1}$ & ${\bf 1}$ & $0$ \\ 
\hline
\end{tabular}
}
\caption{New particles in the type-I seesaw.}
\label{tab:typeI}
\end{table}
}

We begin with the type-I seesaw, arguably the simplest neutrino mass
model.  In this model, the SM particle content is extended with the
addition of $n_N$ generations of right-handed neutrinos $N$, singlets
under the SM gauge group, as shown in Tab.~\ref{tab:typeI}. We will
consider below the most common scenarios, with $n_N = 3$ and $n_N =
2$. The model includes two new Lagrangian terms
\begin{equation}
- \mathcal{L}_{\rm typeI} = y \, H \, \overline{N} \, L 
        + \frac{1}{2} M_N \, \overline{N^c} N + \hc \, ,
\label{typeIlagrangian}
\end{equation}
where we omit flavor indices to simplify the notation. $y$ is a
general $3 \times n_N$ Yukawa matrix while $M_N$ is a $n_N \times n_N$
symmetric mass matrix. The scalar potential of the model is exactly
the same as in the SM. Therefore, symmetry breaking takes place as in
the SM, with the Higgs doublet developing a VEV,
\begin{equation}
\langle H^0 \rangle = \frac{v}{\sqrt{2}} \, .
\end{equation}

After symmetry breaking, the left-handed neutrinos $\nu_L$, the
neutral components of the $L$ lepton doublet, mix with the
right-handed neutrinos $N$. In the basis $(\nu_L\,,\;N^c)$, the
resulting $(3+n_N) \times (3+n_N)$ neutral fermion mass matrix is
given by
\begin{equation}
\mathcal M_{\rm type-I}=\left(\begin{array}{c c} 
0 & m_D^T \\ 
m_D & M_N
\end{array}\right) \, ,
\label{eq:typeImatrix}
\end{equation}
where we have defined $m_D = \frac{1}{\sqrt{2}} \, y \, v$. Under the
assumption $\forall \xi_{ij}\ll 1$, where $\xi = m_D^T
M_N^{-1}$, the mass matrix $\mathcal M_{\rm type-I}$ can be
block-diagonalized to give an effective mass matrix for the $3$ light
neutrinos~\footnote{In models with extra singlet fermions, such as the
  seesaw, there will be non-zero mixing between the active and sterile
  neutrino sectors. This mixing necessarily shows up as non-unitarity
  in the lepton mixing matrix $U$. From the viewpoint of the master
  formula, this corresponds to higher order terms in the seesaw
  expansion $\xi$, which we do not take into account. Since current
  constraints on non-unitarity are of the order of (1-5) percent
  \cite{Escrihuela:2016ube,Blennow:2016jkn,Escrihuela:2019mot}, we do
  not consider this effect numerically very relevant. See
  \cite{Branco:2019avf} for a recent work where these effects are
  addressed.}
\begin{equation}
m_{\rm type-I} = - m_D^T \, {M_N}^{-1} \, m_D 
             = - \frac{v^2}{2} \, y^T \, {M_N}^{-1} \, y\, .
\label{eq:numasstypeI}
\end{equation}
Eq. \eqref{eq:numasstypeI} is shown diagrammatically in
Fig.~\ref{fig:ISS}.  We now compare the type-I seesaw neutrino mass
matrix in Eq. \eqref{eq:numasstypeI} to the general master formula in
Eq. \eqref{eq:master} to establish the following \emph{dictionary}:
\begin{gather}
f = -1 \nonumber \\
n_1 = n_2 = n_N \nonumber \\
y_1 = y_2 = \frac{y}{\sqrt{2}} \nonumber \\
M = \frac{v^2}{2} \, M_N^{-1}
\end{gather}
Furthermore, a symmetric matrix $M$ can be diagonalized by a single
matrix, $V_1 = V_2$, which can be taken to be the identity in this
model, since the right-handed neutrinos can be rotated to their mass
basis without loss of generality. For $n_1 = n_2 = n = r$ the matrices
$X_{1,2,3}$, $\overline W$ and $\overline B$ drop from all the
expressions. We now consider the cases $n_N = 3$ and $n_N = 2$
separately.

\begin{figure}[t]
\centering
\includegraphics[width=0.55\textwidth]{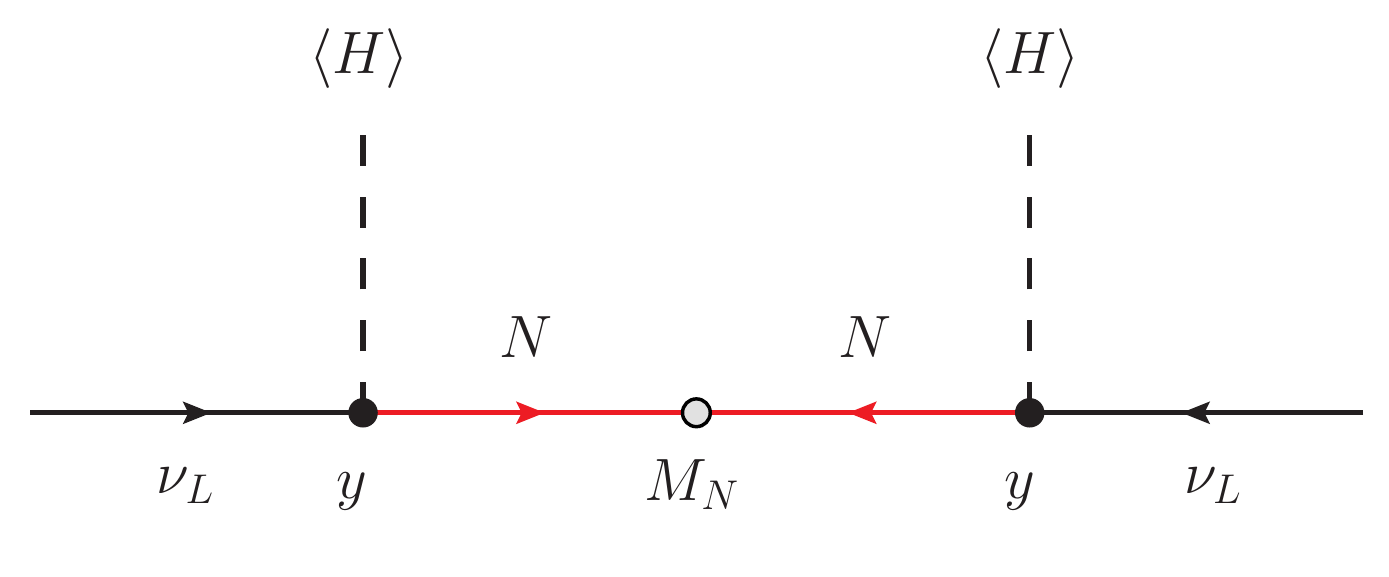}
\caption{Neutrino mass generation in the type-I seesaw.}
\label{fig:typeI}
\end{figure}

\subsubsection{$3$ right-handed neutrinos}

We can now adopt the common choice $r = r_m = 3$, which implies $C_1 =
C_2 = \id_3$. In this case, imposing $y_1 = y_2$ is equivalent to $W^T
W A = B$. Solving this matrix equation leads to $B =
\left(A^T\right)^{-1}$ and allows one to define $R = W \, T = W \, A$,
with $R$ a general $3 \times 3$ orthogonal matrix. Replacing all these
ingredients into Eqs. \eqref{eq:par1} and \eqref{eq:par2} one finds
\begin{equation} \label{eq:CI}
y = \sqrt{2} \, y_1 = \sqrt{2} \, y_2 = i \, \Sigma^{-1/2} \, R \, D_{\sqrt{m}} \, U^\dagger \, ,
\end{equation}
which is nothing but the Casas-Ibarra parametrization for the type-I
seesaw Yukawa matrices. We note that $R$ can be identified with the
usual Casas-Ibarra matrix~\cite{Casas:2001sr}. We conclude that the
Casas-Ibarra parametrization can be regarded as a particular case of
the general master parametrization.

As a final comment, we note that in the type-I seesaw with $3$
generations of right-handed neutrinos, the condition $y_1 = y_2$
implies $18 \, (= 9 \cdot 2)$ real constraints, this is, $\npar_{\rm
  extra} = 18$. Therefore, direct application of the general counting
formula in Eq.~\eqref{eq:counting} leads to $\npar_{\rm free} =
6$. These are the free real parameters contained in the Casas-Ibarra
$R$ matrix, which can be parametrized by means of $3$ complex angles,
see Appendix~\ref{sec:mat}.

\subsubsection{$2$ right-handed neutrinos}

In the type-I seesaw with $2$ generations of right-handed neutrinos
one also obtains the neutrino mass matrix in
Eq.~\eqref{eq:numasstypeI}, but with $n_1 = n_2 = n = r =
2$. Moreover, it is well known that in this case one induces only two
non-vanishing neutrino mass eigenvalues, and hence $r_m = 2$ and the
model belongs either to the $(2,2)_a$ case or to the $(2,2)_b$ case.
One can now follow a similar approach as for the $3$ generation
model. In the $2$ generation version, imposing $y_1 = y_2$ is
equivalent to $W^T W A = B \leftrightarrow T^T \, W^T \, W \, T \, C_1
= C_1 \, C_2 + K C_1$. Replacing the expressions for $C_1$ and $C_2$
in the $(2,2)_b$ case, one can easily find that this matrix equation
leads to a contradiction. In case of neutrino NH this is found by
comparing the elements $(1,2)$ and $(1,3)$, whereas in case of IH by
comparing the elements $(1,1)$ and $(1,2)$. Therefore, we discard this
scenario.  Solving the matrix equation (decomposing it by elements) in
the $(2,2)_a$ case, leads to $z_1=z_2=0$, $K=0$ and $R = W \, T$, with
$R$ a general $2 \times 2$ orthogonal matrix that can be parametrized
by one complex angle. In summary, replacing all these ingredients into
Eqs. \eqref{eq:par1} and \eqref{eq:par2} one finds
\begin{equation} \label{eq:CI2}
y = \sqrt{2} \, y_1 = \sqrt{2} \, y_2 = i \, \Sigma^{-1/2} \, R \, 
\left( \begin{array}{ccc} 0 & \sqrt{m_2} & 0 \\ 0 & 0 & \sqrt{m_3} \end{array} 
\right) \, P \, U^\dagger \, ,
\end{equation}
where $P = \id_3$ in case of NH and $P = P_{13}$ and in case of IH,
see Eqs.~\eqref{eq:Pdef} and \eqref{eq:P13}. In case of IH one should
also rename $m_3 \to m_1$. The result in Eq.~\eqref{eq:CI2} agrees
perfectly with~\cite{Ibarra:2003up}.

\subsection{The inverse seesaw}
\label{subsec:inverse}

{
\renewcommand{\arraystretch}{1.4}
\begin{table}
\centering
{\setlength{\tabcolsep}{0.5em}
\begin{tabular}{| c c c c c c |}
\hline  
 & spin & generations & $\mathrm{SU(3)}_c$ & $\mathrm{SU(2)}_L$ 
 & $\mathrm{U(1)}_Y$ \\
\hline
\hline    
$N$ & 1/2 & 3 & ${\bf 1}$ & ${\bf 1}$ & $0$ \\ 
$S$ & 1/2 & 3 & ${\bf 1}$ & ${\bf 1}$ & $0$ \\ 
\hline
\end{tabular}
}
\caption{New particles in the inverse seesaw.}
\label{tab:ISS}
\end{table}
}

\begin{figure}[t]
\centering
\includegraphics[width=0.55\textwidth]{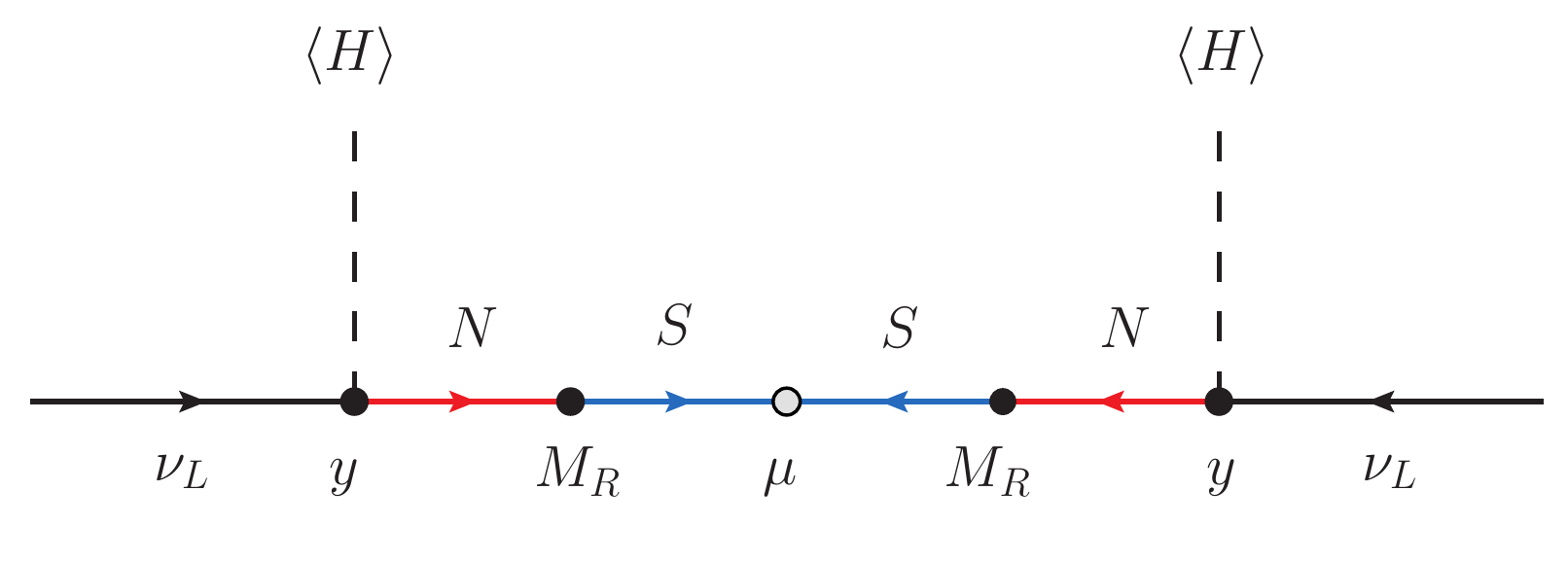}
\caption{Neutrino mass generation in the inverse seesaw.}
\label{fig:ISS}
\end{figure}

We now consider the inverse seesaw \cite{Mohapatra:1986bd}, an example
model in which the matrix $M$ is actually the product of several
matrices.  In the inverse seesaw, the SM particle content is extended
with the addition of $3$ generations of right-handed neutrinos $N$ and
$3$ generations of singlet fermions $S$, both with lepton number $+1$,
as summarized in Tab.~\ref{tab:ISS}.~\footnote{See
  \cite{Malinsky:2009df,Gavela:2009cd,Abada:2014vea} for more minimal
  realizations of the inverse seesaw.}  The Lagrangian is assumed to
contain the following terms involving these fields
\begin{equation}
- \mathcal{L}_{\rm ISS} = y \, H \, \overline{N} \, L 
 + M_R \, \overline{N} \, S + \frac{1}{2} \mu \, \overline{S^c} S + \hc \, ,
\label{ISSlagrangian}
\end{equation}
 where we omit flavor indices to simplify the notation. $y$ is a
general $3 \times 3$ Yukawa matrix, $M_R$ is an arbitrary complex $3
\times 3$ mass matrix while $\mu$ is a $3 \times 3$ complex symmetric
mass matrix. Again, the scalar potential and symmetry breaking pattern
of the model is the same as in the SM. After symmetry breaking, the
left-handed neutrinos $\nu_L$ mix with the $N$ and $S$ singlet
fermions. In the basis $(\nu_L\,,\;N^c\,,\;S)$, the resulting $9\times
9$ neutral fermion mass matrix is given by
\begin{equation}
\mathcal M_{\rm ISS}=\left(\begin{array}{c c c} 
0 & m_D^T & 0 \\ 
m_D & 0 & M_R \\ 
0 & M_R^T & \mu 
\end{array}\right) \, .
\label{eq:ISSmatrix}
\end{equation}
We note that in the absence of the $\mu$ term, the matrix in
Eq. \eqref{eq:ISSmatrix} would have a Dirac structure and lead to
three massless states. In fact, $\mu$ violates lepton number by two
units and can be taken naturally small, in the sense of 't
Hooft~\cite{tHooft:1979rat}, since the limit $\mu \to 0$ restores
lepton number and increases the symmetry of the model. Under the
assumption $\mu \ll m_D \ll M_R$, the mass matrix $\mathcal M_{\rm
  ISS}$ can be block-diagonalized to give an effective mass matrix for
the $3$ light neutrinos~\cite{GonzalezGarcia:1988rw}
\begin{equation}
m_{\rm ISS} = m_D^T \, {M_R^T}^{-1} \, \mu \, M_R^{-1} \, m_D 
   = \frac{v^2}{2} \, y^T \, {M_R^T}^{-1} \, \mu \, M_R^{-1} \, y\, .
\label{eq:numassISS}
\end{equation}
Eq. \eqref{eq:numassISS} is shown diagrammatically in
Fig.~\ref{fig:ISS}.  Again, we can compare the inverse seesaw neutrino
mass matrix in Eq. \eqref{eq:numassISS} to the general master formula
in Eq. \eqref{eq:master} and establish a \emph{dictionary}:~\footnote{We point out that this is just one possible dictionary. For instance, one could include the $\frac{v^2}{2}$ factor in the definition of $f$ and modify $M$ accordingly.}
\begin{gather}
f = 1 \nonumber \\
n_1 = n_2 = 3 \nonumber \\
y_1 = y_2 = \frac{y}{\sqrt{2}} \nonumber \\
M = \frac{v^2}{2} \, {M_R^T}^{-1} \, \mu \, M_R^{-1}
\end{gather}
This identification clearly shows that one can make use of an adapted
Casas-Ibarra parametrization for the inverse seesaw
\cite{Deppisch:2004fa}.

However, compared to the simpler type-I seesaw, discussed above, here
$M$ can not be taken taken to be diagonal automatically and $V_1=V_2$
become physical. (Note that the two rotation matrices are still equal,
since $M$ is a complex symmetric matrix in the inverse seesaw.)  The
reason for this is straightforward: $M$ contains the two matrices
$M_R$ and $\mu$. If $y$ is taken arbitrary, we can still use field 
redefinitions for $N$ and $S$ to choose either $M_R$ {\em or} $\mu$
diagonal, but not both at the same time.

\subsection{The scotogenic model}
\label{subsec:scotogenic}

This example illustrates the use of the master parametrization in a
model with loop induced neutrino masses. As we will show below, the
radiative origin of neutrino masses does not alter the application of
the master parametrization.

The \emph{scotogenic model}~\cite{Ma:2006km} extends the SM particle
content with three generations of the singlet fermions $N$ and the
$SU(2)_L$ doublet scalar $\eta$. In addition, a $\mathbb{Z}_2$
symmetry is imposed, under which the new particles are odd while the
SM ones are assumed to be even.  The quantum numbers of the new
particles in the scotogenic model are given in
Table~\ref{tab:scotogenic}.

In addition to the canonical kinetic term, the Lagrangian contains the
following terms involving the singlet fermions, 
\begin{equation}
- \mathcal{L}_{\rm SC} = y \, \eta \, \overline{N} \, L
+ \frac{M_N}{2} \, \overline{N^c} \, N
+ \hc \, ,
\end{equation}
where we omit flavor indices for the sake of clarity. Here $M_N$ is a
$3 \times 3$ symmetric matrix with dimensions of mass which can be
taken to be diagonal without loss of generality. The matrix of Yukawa
couplings, $y$, is an arbitrary $3 \times 3$ complex matrix. The
scalar potential of the model is given by
\begin{align}
\mathcal{V}_{\rm SC} =&\:
m_H^2 H^\dag H + m_\eta^2 \eta^\dag \eta+
\frac{\lambda_1}{2}\left(H^\dag H \right)^2+
\frac{\lambda_2}{2}\left(\eta^\dag \eta\right)^2+
\lambda_3\left(H^\dag H\right)\left(\eta^\dag\eta\right)\nonumber\\
&+
\lambda_4\left(H^\dag\eta\right)\left(\eta^\dag H\right)+
\left[ \frac{\lambda_5}{2} \left(H^\dag\eta\right)^2+
\frac{\lambda_5^\ast}{2} \left(\eta^\dag H\right)^2\right] \, .
\end{align}
All parameters in the scalar potential are real, with the exception of
the $\lambda_5$ quartic parameter, which can be complex. In the
scotogenic model, the $\mathbb{Z}_2$ parity is assumed to be preserved
after symmetry breaking. This is guaranteed by choosing a set of
parameters that leads to a vacuum with
\begin{equation}
\langle H^0 \rangle = \frac{v}{\sqrt{2}} \quad , \quad \langle \eta^0 \rangle = 0 \, .
\end{equation}
After electroweak symmetry breaking, the masses of the charged
component $\eta^+$ and neutral component
$\eta^0=(\eta_R+i \,\eta_I)/\sqrt{2}$ are split to
\begin{eqnarray}
m_{\eta^+}^2&=&m_\eta^2+\lambda_3\langle H^0\rangle^2 \, , \\
m_R^2&=&m_{\eta}^2+\left(\lambda_3+\lambda_4+\lambda_5\right)\langle H^0\rangle^2 \, ,\\
m_I^2&=&m_{\eta}^2+\left(\lambda_3+\lambda_4-\lambda_5\right)\langle H^0\rangle^2 \, . 
\end{eqnarray}
We note that the mass difference between $\eta_R$ and $\eta_I$ (the
CP-even and CP-odd components of the neutral $\eta^0$, respectively)
is controlled by the $\lambda_5$ coupling since
$m_R^2-m_I^2=2\lambda_5\langle H^0\rangle^2$. This will be relevant
for the generation of non-vanishing neutrino masses in this model.

One of the most attractive features of the scotogenic model is the
presence of a dark matter candidate. Indeed, the conservation of the
$\mathbb{Z}_2$ symmetry implies that the lightest state charged under
this parity is completely stable and, in principle, can serve as a
good dark matter candidate. This role can be played by the lightest
singlet fermion ($N_1$) or by the neutral component of the
\emph{inert} $\eta$ doublet ($\eta_R$ or $\eta_I$).

{
\renewcommand{\arraystretch}{1.4}
\begin{table}
\centering
{\setlength{\tabcolsep}{0.5em}
\begin{tabular}{| c c c c c c c |}
\hline  
 & spin & generations & $\mathrm{SU(3)}_c$ & $\mathrm{SU(2)}_L$ 
        & $\mathrm{U(1)}_Y$ & $\mathbb{Z}_2$ \\
\hline
\hline    
$\eta$ & 0 & 1 & ${\bf 1}$ & ${\bf 2}$ & $1/2$ & $-$ \\
\hline
\hline    
$N$ & 1/2 & 3 & ${\bf 1}$ & ${\bf 1}$ & $0$ & $-$ \\ 
\hline
\end{tabular}
}
\caption{New particles in the scotogenic model.}
\label{tab:scotogenic}
\end{table}
}

\begin{figure}[t]
\centering
\includegraphics[width=0.5\textwidth]{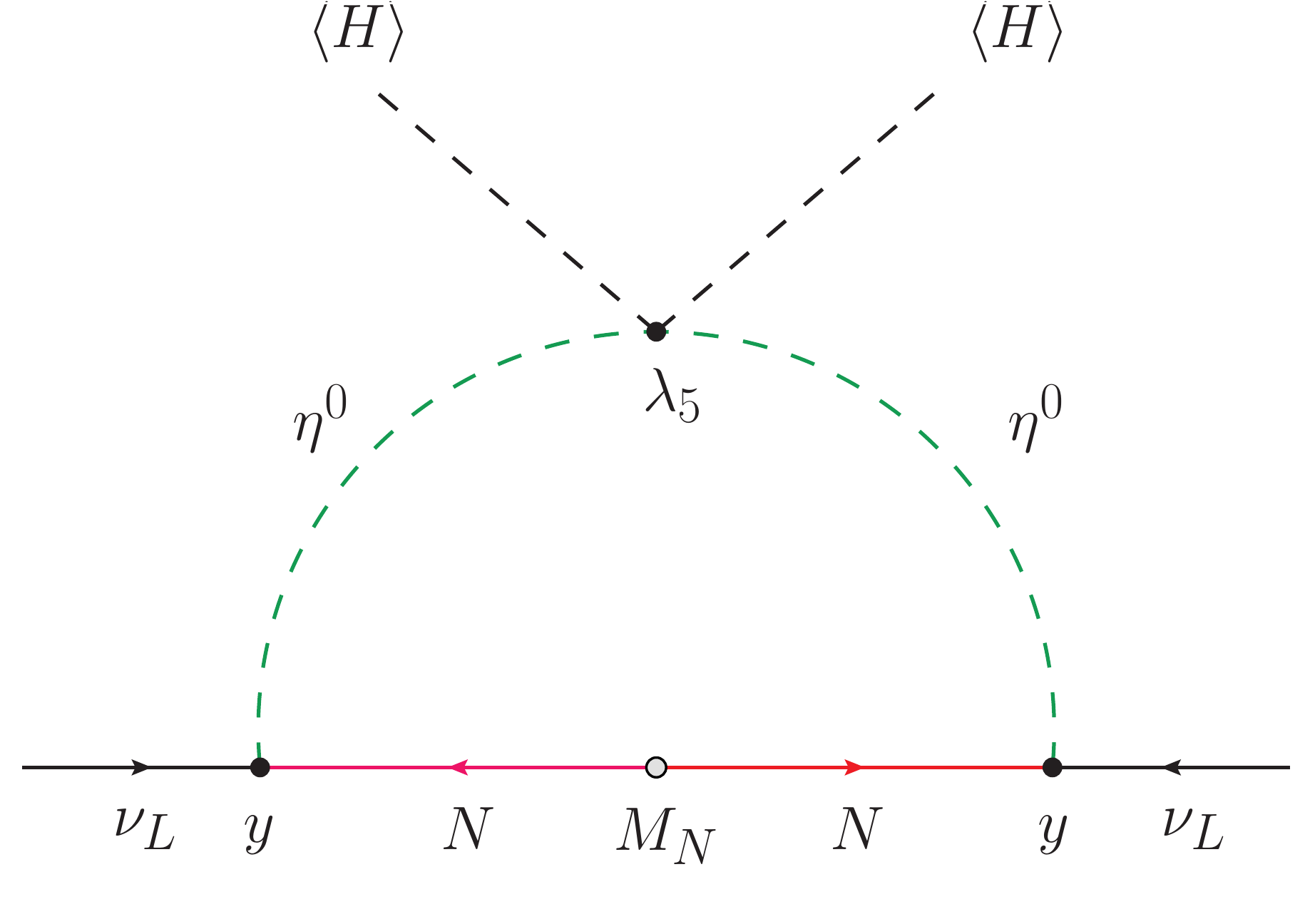}
\caption{Neutrino mass generation in the scotogenic
  model.}
\label{fig:SC}
\end{figure}

We now move to the discussion of neutrino masses. First, we note that
the singlet fermions do not couple to the SM Higgs doublet due to the
$\mathbb{Z}_2$ discrete symmetry while $\langle \eta^0 \rangle = 0$
prevents the $\eta \, \overline{N} \, L$ Yukawa term from inducing a
Dirac mass term for the neutrinos. Therefore, neutrino masses vanish
at tree-level but get induced at the 1-loop level, as shown in
Fig. \ref{fig:SC}. The resulting $3 \times 3$ Majorana
neutrino mass matrix is given by
\begin{align}
\left(m_{\rm SC}\right)_{\alpha\beta} &=
\sum_{i=1}^3\frac{y_{i\alpha}y_{i\beta}}{2(4\pi)^2} M_{N_i}
\left[\frac{m_R^2}{m_R^2-M_{N_i}^2}\log\left(\frac{m_R^2}{M_{N_i}^2}\right)
-\frac{m_I^2}{m_I^2-M_{N_i}^2}\log\left(\frac{m_I^2}{M_{N_i}^2}\right)\right] \, ,
\label{eq:numassSC}
\\ 
 & \equiv \frac{1}{32 \pi^2} \left( y^T \, {\widehat M} \, y \right)_{\alpha \beta} \, ,
\label{eq:numassSC1a}
\end{align}
with the diagonal matrix ${\widehat M}$ with entries
\begin{equation}
{\widehat M}_{ii} =M_{N_i}
\left[\frac{m_R^2}{m_R^2-M_{N_i}^2}\log\left(\frac{m_R^2}{M_{N_i}^2}\right)
-\frac{m_I^2}{m_I^2-M_{N_i}^2}\log\left(\frac{m_I^2}{M_{N_i}^2}\right)\right] \, ,  
\label{eq:Lam1}
\end{equation}
A simplified expression can be obtained when $m_R^2 \approx m_I^2
\equiv m_0^2$ (or, equivalently, $\lambda_5\ll 1$). In this case,
Eq. \eqref{eq:numassSC} reduces to~\footnote{We note that
  $\lambda_5 \ll 1$ is a natural choice in the sense of 't
  Hooft~\cite{tHooft:1979rat}, since the limit $\lambda_5 \to 0$
  increases the symmetry of the model by restoring lepton number.}
\begin{align} \label{eq:numassSC2}
\left(m_{\rm SC}\right)_{\alpha\beta} & \simeq
\sum_{i=1}^3\frac{\lambda_5 y_{i\alpha}y_{i\beta}\langle H^0\rangle^2}
{(4\pi)^2 M_{N_i}}
\left[\frac{M_{N_i}^2}{m_{0}^2-M_{N_i}^2}
+\frac{M_{N_i}^4}{\left(m_{0}^2-M_{N_i}^2\right)^2}
\log\left(\frac{M_{N_i}^2}{m_0^2}\right)\right] \nonumber \\
& \equiv \frac{\lambda_5}{16 \pi^2} \left( y^T \Lambda \, y \right)_{\alpha \beta} \, ,
\end{align}
where we have defined $\Lambda = \text{diag}\left(
\Lambda_1,\Lambda_2,\Lambda_3 \right)$, with
\begin{equation}
\Lambda_i = \frac{\langle H^0\rangle^2}{M_{N_i}}
\left[\frac{M_{N_i}^2}{m_{0}^2-M_{N_i}^2}
  +\frac{M_{N_i}^4}{\left(m_{0}^2-M_{N_i}^2\right)^2}
  \log\left(\frac{M_{N_i}^2}{m_0^2}\right)\right] \, .
\end{equation}
Eq. \eqref{eq:numassSC1a} and the last equality of
Eq. \eqref{eq:numassSC2} clearly shows that the Yukawa matrix $y$ can
be written using an adapted Casas-Ibarra parametrization
\cite{Toma:2013zsa}. In fact, direct comparison to the master formula
in Eq. \eqref{eq:master} allows one to identify
\begin{gather}
f = \frac{\lambda_5}{16 \pi^2} \nonumber \\
n_1 = n_2 = 3 \nonumber \\
y_1 = y_2 = \frac{y}{\sqrt{2}} \nonumber \\
M = \Lambda
\end{gather}
in the scotogenic model.

\subsection{The linear seesaw}
\label{subsec:linear}

The full power of the master parametrization is better illustrated
with an application to the linear
seesaw~\cite{Akhmedov:1995ip,Akhmedov:1995vm}, which provides a
well-known example of a neutrino mass formula with $y_1 \ne y_2$.

{
\renewcommand{\arraystretch}{1.4}
\begin{table}
\centering
{\setlength{\tabcolsep}{0.5em}
\begin{tabular}{| c c c c c c |}
\hline  
 & spin & generations & $\mathrm{SU(3)}_c$ & $\mathrm{SU(2)}_L$ 
        & $\mathrm{U(1)}_Y$ \\
\hline
\hline    
$N$ & 1/2 & 3 & ${\bf 1}$ & ${\bf 1}$ & $0$ \\ 
$S$ & 1/2 & 3 & ${\bf 1}$ & ${\bf 1}$ & $0$ \\ 
\hline
\end{tabular}
}
\caption{New particles in the linear seesaw.}
\label{tab:LSS}
\end{table}
}

\begin{figure}[t]
\centering
\includegraphics[width=0.45\textwidth]{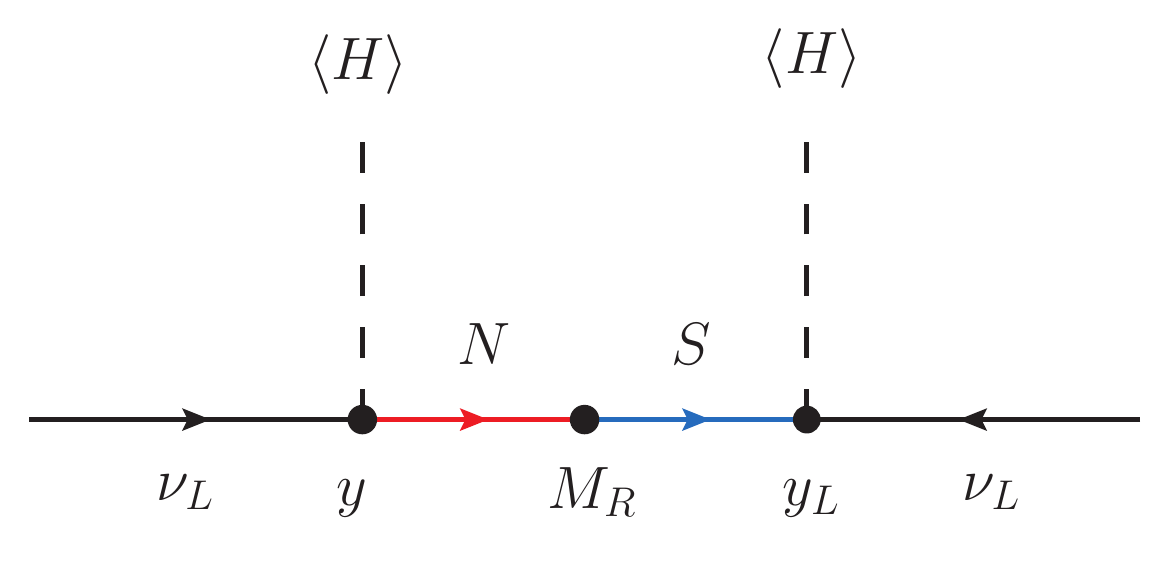}
\caption{Neutrino mass generation in the linear seesaw.}
\label{fig:LSS}
\end{figure}

Originally introduced in the context of left-right symmetric
models~\cite{Akhmedov:1995ip,Akhmedov:1995vm}, this mechanism has also
been shown to arise naturally in $\mathrm{SO(10)}$ unified
theories~\cite{Barr:2003nn,Malinsky:2005bi}. The particle content of
the model is the same as in the inverse seesaw, as shown in
Tab.~\ref{tab:LSS}.  The Lagrangian is assumed to contain the
following terms
\begin{equation}
- \mathcal{L}_{\rm LSS} = y \, H \, \overline{N} \, L 
  + M_R \, \overline{N} \, S + y_L \, H \, \overline{L^c} \, S + \hc \, ,
\label{LSSlagrangian}
\end{equation}
where again we omit flavor indices to simplify the notation. As in the
inverse seesaw, $y$ is a general $3 \times 3$ Yukawa matrix and $M_R$
is a $3 \times 3$ complex mass matrix. In addition, $y_L$ is a general
$3 \times 3$ Yukawa matrix, with $y_L \ne y$ in general. Therefore,
the linear seesaw model features $y_1 \ne y_2$. The scalar potential
and symmetry breaking pattern of the model is the same as in the
SM. In the basis $(\nu_L\,,\;N^c\,,\;S)$, the resulting $9\times 9$
neutral fermion mass matrix obtained after electroweak symmetry
breaking takes the form
\begin{equation}
\mathcal M_{\rm LSS}=\left(\begin{array}{c c c} 
0 & m_D^T & M_L \\ 
m_D & 0 & M_R \\ 
M_L^T & M_R^T & 0 
\end{array}\right) \, ,
\label{eq:LSSmatrix}
\end{equation}
where $M_L = \frac{1}{\sqrt{2}} \, y_L v$. We note that in the
presence of $M_L$, lepton number is broken in two units. Assuming
$m_D, M_{L} \ll M_{R}$, the mass matrix for the $3$ light neutrinos is
given by
\begin{equation}
m_{\rm LSS} = M_L \, M_R^{-1} \, m_D + m_D^T \, {M_R^T}^{-1} \, M_L^T 
   = \frac{v^2}{2} \, \left( y_L \, M_R^{-1} \, y
    + y^T \, {M_R^T}^{-1} \, y_L^T \right) \, .
\label{eq:numassLSS}
\end{equation}
Eq.~\eqref{eq:numassLSS} is shown diagrammatically in
Fig.~\ref{fig:LSS} (without the transposed 2nd term). We see that the
resulting expression for the light neutrino mass matrix is linear in
$y$ (or, equivalently, in $m_D$), hence the origin of the name
\emph{linear seesaw}. As usual, we now compare the linear seesaw
neutrino mass matrix in Eq.~\eqref{eq:numassLSS} to the general master
formula in Eq.~\eqref{eq:master}. By doing so one finds the following
\emph{dictionary}:
\begin{gather}
f = 1 \nonumber \\
n_1 = n_2 = 3 \nonumber \\
y_1 = y_L^T \nonumber \\
y_2 = y \nonumber \\
M = \frac{v^2}{2} \, M_R^{-1}
\end{gather}
We emphasize again that one cannot make use of the standard
Casas-Ibarra parametrization in the linear seesaw model due to $y \ne
y_L^T$ (a particular example of the general case $y_1 \ne y_2$). In this
case one must necessarily make use of the full master parametrization.

\section{Models with extra symmetries and restrictions}
\label{sec:extra}

We now discuss Majorana neutrino mass models which follow the
structure of Eq.~\eqref{eq:master}, but the master parametrization
may become either not direct, impractical or
useless. These ``exceptional'' cases are simply those for which $y_1$
and $y_2$ are not completely free parameters. Based on the type of
restrictions, the $y_1$ and $y_2$ Yukawa matrices must follow, one can
identify four categories:
\begin{enumerate} [label=(\bf \roman*)]
\item {\bf Identity models}: $y_1 = y_2 = \id$. This is the case of the
  seesaw type-II and similar models.
\item {\bf Symmetric models}: $y_1 = y_1^T$ and/or $y_2 = y_2^T$. This
  is the case in many models with an underlying left-right symmetry.
\item {\bf Antisymmetric models}: $y_1 = -y_1^T$ and/or $y_2 =
  -y_2^T$. This scenario takes place in models including the charged
  scalar $s$, which transforms as $({\bf 1}, {\bf 1}, 1)$ under the SM
  gauge symmetry, due to the presence of the antisymmetric $\rm
  SU(2)_L$ contraction in the $s \, \overline{L^c} \, L$ Yukawa
  term. Two well known examples of such scenario are the Zee ($y_1 = -
  y_1^T$) and Zee-Babu ($\sqrt{2} \,y_1 = \sqrt{2} \,y_2 = y$, with $y
  = - y^T$) models.
\item {\bf Flavored models}: specific textures in $y_1$ and $y_2$. This
  would be the case of models with flavor symmetries. Models with
  conditions on the $y_1$ and $y_2$ Yukawa matrices not included in
  the previous cases can be generically included here.
\end{enumerate}
As discussed next, case (i) is trivial, whereas case (ii) needs only a
slight modification of our procedure. Only cases (iii) and (iv) are
not so easily solved and require an in-depth discussion. \\

Identity models, those with $y_1 = y_2 = \id$, are trivially
addressed. For instance, let us consider the type-II
seesaw~\cite{Mohapatra:1980yp,Schechter:1980gr}. This model extends
the SM particle content with the $\rm SU(2)_L$ triplet scalar $\Delta$
with hypercharge $Y_\Delta = 1$. The inclusion of this field allows us
to write the Yukawa term $Y_\Delta \, \overline{L^c} \, \Delta \, L$
which, after the neutral component of $\Delta$ acquires a VEV,
$v_\Delta$, induces Majorana masses for the neutrinos, with their mass
matrix given by $m = Y_\Delta \, v_\Delta$. It is clear that this
model can also be described by means of the master formula, with the
dictionary simply given by
\begin{gather}
f = 1 \nonumber \\
n_1 = n_2 = 3 \nonumber \\
y_1 = y_2 = \id_3 \nonumber \\
M = Y_\Delta \, v_\Delta
\end{gather}
Even though the master formula also includes models in this category,
they do not require a parametrization for the Yukawa matrices. Note
that the neutrino mixing matrix is simply given by the diagonalization
matrix of $Y_{\Delta}$. 

In what concerns symmetric models, a simple yet elegant solution when
$y_1 = y_2$ was given in \cite{Anamiati:2016uxp}. We proceed to
reproduce it here. Let us consider a fully symmetric type-I seesaw
neutrino mass matrix with $\sqrt{2} \, y_1 = \sqrt{2} \, y_2 = y =
y^T$. The master formula reduces to $m = y^T M y \equiv y M y$ and the
master parametrization to a Casas-Ibarra parametrization, see
Eq.~\eqref{eq:CI}. $M$ must be a symmetric matrix in this case, and
then it can be brought to a diagonal form with just a single matrix $V$,
\begin{equation}
M = V^T \, \widehat{\Sigma} \, V \, ,
\end{equation}
and the Casas-Ibarra parametrization reads
\begin{equation} \label{eq:CIsym}
y = i \, V^\dagger \, \Sigma^{-1/2} \, R \, D_{\sqrt{m}} \, U^\dagger \, ,
\end{equation}
with $R$ a complex orthogonal $3 \times 3$ matrix. This equation can be
trivially rewritten as
\begin{equation} \label{eq:CIsym2}
Y = V \, y = i \, \Sigma^{-1/2} \, R \, D_{\sqrt{m}} \, U^\dagger \, .
\end{equation}
This shows that the matrix $Y$ can be obtained by applying a standard
Casas-Ibarra parametrization. The key now is to be able to decompose
it as the product of the unitary matrix $V$ and the symmetric matrix
$y$. In order to do that we first apply a singular-value decomposition,
\begin{equation}
Y = W_1^T \, \widehat Y \, W_2 \, ,
\end{equation}
where $W_1$ and $W_2$ are two unitary matrices and $\widehat Y$ is a
diagonal matrix containing the (real and non-negative) singular values
of $Y$. We can now insert $W_2^\ast W_2^T = \id_3$ to obtain
\begin{equation}
Y = W_1^T \, \widehat Y \, W_2 = \left( W_1^T \, W_2^\ast \right) \, \left( W_2^T \, \widehat Y \, W_2 \right) \equiv \widetilde V \, \widetilde y \, ,
\end{equation}
where we have identified the unitary matrix $\widetilde V = W_1^T \,
W_2^\ast$ and the symmetric matrix $\widetilde y = W_2^T \, \widehat Y
\, W_2$. As explained in \cite{Anamiati:2016uxp}, $\widetilde V$ and
$\widetilde y$ are not unique, simply because the singular-value
decomposition is not unique. One can always define
\begin{align}
W_1^\prime &= D_\phi \, W_1 \, , \label{eq:Wphi} \\
W_2^\prime &= D_\phi^{-1} \, W_2 \, , \label{eq:Vphi}
\end{align}
with
\begin{equation} \label{eq:Dphi}
D_\phi = \text{diag} \left(e^{i \phi_1}, e^{i \phi_2}, e^{i \phi_3} \right)
\end{equation}
a diagonal phase matrix, such that $Y = W_1^{\prime \, T} \, \widehat
Y \, W_2^\prime$ as well.~\footnote{In general, the singular-value
  decomposition is unique up to arbitrary unitary transformations
  applied uniformly to the column vectors of both $W_1$ and $W_2$
  spanning the subspaces of each singular value, and up to arbitrary
  unitary transformations on vectors of $W_1$ and $W_2$ spanning the
  kernel and cokernel, respectively, of $Y$. This well-known fact is
  reflected, for example, in the freedom in the determination of
  eigenvectors for a set of degenerate eigenvalues.}  These three
phases must be taken into account in the factorization of $Y$ as the
product of a unitary matrix and a symmetric matrix. We then make the
identification
\begin{align}
V &= W_1^T \, D_\phi^{-1} \, W_2^\ast \, , \\
y &= W_2^T \, D_\phi \, \widehat Y \, W_2\, ,
\label{eq:decom}
\end{align}
which preserves $Y = V \, y$ and the symmetric nature of $y$. In summary,
when both Yukawa matrices are equal and symmetric, one can use the
standard Casas-Ibarra parametrization for $Y$ and finally find $y$ by
means of the decomposition in Eq.~\eqref{eq:decom}.

Finally, we come to case (iii), models with antisymmetric Yukawa
matrices. We first consider the scenario with one antisymmetric Yukawa
coupling, $y_1 = - y_1^T$, with general $y_2$. The most popular model
of this class is the Zee model~\cite{Zee:1980ai}, discussed in
Sec.~\ref{subsec:zee}. As in the general case, both Yukawa matrices,
$y_1$ and $y_2$, can be written using the master parametrization in
Eqs.~\eqref{eq:par1} and \eqref{eq:par2}. However, the antisymmetry of
$y_1$ implies some non-trivial conditions on the matrices $W$ and $T$,
as well as on $m$ and $M$. Therefore, the \textit{input} matrices $m$
and $M$ can no longer be arbitrary, but are indeed forced to follow
some relations if the master formula in Eq.~\eqref{eq:master} is to be
satisfied. More details about this scenario with one antisymmetric
Yukawa coupling can be found in Appendix~\ref{sec:proof_antisym1}. Now
we turn to the special case of equal and antisymmetric Yukawa
matrices, $\sqrt{2} \, y_1=\sqrt{2} \,y_2=y=-y^T$. The Zee-Babu model
\cite{Cheng:1980qt,Zee:1985id,Babu:1988ki}, presented in detail in
Sec.~\ref{subsec:zeebabu}, is the most popular model of this class. In
this scenario one necessarily has $n_1=n_2=3$, $V_1=V_2 \equiv V$ and
$r=r_m=2$. The master formula reduces to $m = y^T M y = - y M y$ and
the master parametrization to a modified Casas-Ibarra
parametrization. In case of $n=3$ one finds
\begin{equation} \label{eq:CIantisym}
y = \sqrt{2} \, y_1 = \sqrt{2} \, y_2 = i \, V^\dagger \, \Sigma^{-1/2} \, R \, C_1 \, \bar{D}_{\sqrt{m}} \, U^\dagger \, ,
\end{equation}
with $C_1$ given in Eq.~\eqref{eq:C1C222a}, in this case fixing
$z_1=z_2=0$, and $R$ a $3\times 2$ Casas-Ibarra matrix such that $R^T
R = \id_2$. However, the parametrization for the $y$ matrix in
Eq.~\eqref{eq:CIantisym} is not sufficient to guarantee the
antisymmetry of the $y$ Yukawa matrix. Many additional restrictions
must be taken into account. In fact, the equality $\sqrt{2}
\,y_1=\sqrt{2} \,y_2=y=-y^T$ implies 12 (real) conditions. Since the
number of real free parameters in this scenario is 6, the system is
overconstrained. This has two implications. First, in contrast to the
general case, $R$ must take a very specific form. And second, the
parameters in $m$ and $M$ are not free anymore, but they are indeed
forced to follow 6 real conditions: one vanishing neutrino mass
eigenvalue, one vanishing Majorana phase and two (complex) non-trivial
conditions. For the proof and more details about this special case we
refer to Appendix~\ref{sec:proof_antisym2}. 

Let us also comment on alternative approaches in case of antisymmetric
Yukawa couplings. First, in models in which $M$ is a product of more
than one matrix, it may be more practical to solve for (one of) the
\textit{inner} Yukawa couplings, instead of $y_1$ or $y_2$. And
second, we are discussing a master parametrization which we later
particularize to specific models. This approach is completely general
and can be used for any Majorana neutrino mass model. However, in some
particular cases there might be a simpler and more direct
approach. For instance, a parametrization for the antisymmetric
scenario with $\sqrt{2} \, y_1=\sqrt{2} \,y_2=y=-y^T$ was presented in
\cite{Babu:2002uu}. The antisymmetry of the $y$ matrix implies that
\begin{equation}
v_0 = \begin{pmatrix}
y_{23} \\
- y_{13} \\
y_{12}
\end{pmatrix}
\end{equation}
is an eigenvector of $y$ with null eigenvalue. Since $m = - y M y$,
$v_0$ is also eigenvector of $m$ and we can write
\begin{equation} \label{eq:eigen0}
m \, v_0 = 0 \Leftrightarrow U^\ast \, D_m \, U^\dagger \, v_0 = 0 \Leftrightarrow D_m \, U^\dagger \, v_0 = 0 \, .
\end{equation}
This equation can be solved analytically to determine two of the
components of $v_0$ in terms of the third and the neutrino masses and
mixing angles contained in $D_m$ and $U$. Furthermore, as explained
above, the matrix $M$ is not free in this special case. The conditions
on its entries can be derived by replacing the form for $y$ obtained
with Eq.~\eqref{eq:eigen0} into $m = U^\ast \, D_m \, U^\dagger = - y
M y$. Out of the six equations, only three are independent. Therefore,
one can obtain three $M$ entries in terms of the remaining
parameters. For instance, one can choose to solve the equations for
$M_{22}$, $M_{23}$ and $M_{33}$. This solution has been found to be
very convenient for phenomenological studies
\cite{Herrero-Garcia:2014hfa}. Nevertheless, we emphasize again that
our focus is on the generality of our approach, while this type of
solutions can only be applied to very specific scenarios.

Finally, the number of possible restrictions in flavored models is
enormous and a systematic exploration is not feasible. For this
reason, we will not discuss them here, although we note that the
master parametrization might provide a powerful analytical tool for
the treatment of these special cases. We also point out that in some
models the charged lepton mass matrix is not diagonal in the flavor
basis. Instead, the mass and flavor bases are related by
\begin{equation}
\widehat m_e = U_e^\dagger \, m_e \, V_e \, ,
\end{equation}
where $m_e$ and $\widehat m_e$ are the charged lepton mass matrix in
the flavor and mass bases, respectively, and $U_e$ and $V_e$ are two
$3 \times 3$ unitary matrices. This would introduce an additional
unitary matrix in the master parametrization, replacing $U^\dagger$ in
Eqs.~\eqref{eq:par1} and \eqref{eq:par2} by $U^\dagger \, U_e^\dagger$. \\

We now present two models of type (iii), the Zee and Zee-Babu
models. They constitute well-known examples of models with
antisymmetric Yukawa couplings.

\subsection{The Zee model}
\label{subsec:zee}

{
\renewcommand{\arraystretch}{1.4}
\begin{table}
\centering
{\setlength{\tabcolsep}{0.5em}
\begin{tabular}{| c c c c c c |}
\hline  
 & spin & generations & $\mathrm{SU(3)}_c$ & $\mathrm{SU(2)}_L$ & $\mathrm{U(1)}_Y$ \\
\hline
\hline    
$\phi$ & 0 & 1 & ${\bf 1}$ & ${\bf 2}$ & $1/2$ \\ 
$s$ & 0 & 1 & ${\bf 1}$ & ${\bf 1}$ & $1$ \\ 
\hline
\end{tabular}
}
\caption{New particles in the Zee model.}
\label{tab:Z}
\end{table}
}

\begin{figure}[t]
\centering
\includegraphics[width=0.45\textwidth]{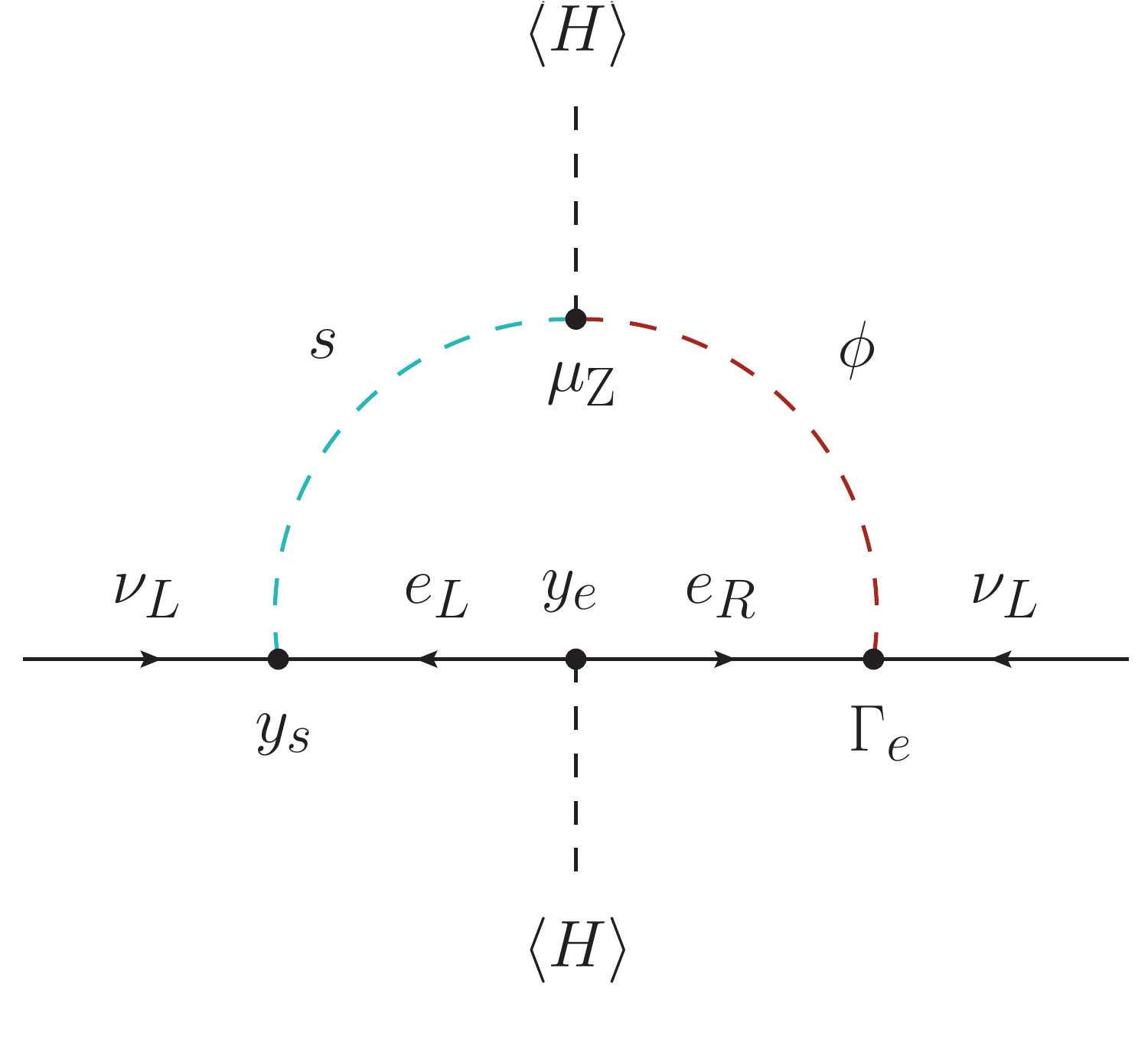}
\caption{Neutrino mass generation in the Zee model.}
\label{fig:Z}
\end{figure}

The Zee model~\cite{Zee:1980ai} constitutes a very simple scenario
beyond the SM leading to radiative neutrino masses. The particle
content of the SM is extended to include a second Higgs doublet,
$\phi$, and the $\rm SU(2)_L$ singlet scalars $s$, with hypercharge
$+1$. Therefore, the Zee model can be regarded as an extension of the
general Two Higgs Doublet Model (THDM) by a charged scalar. As we will
see below, the presence of this singly-charged scalar has a strong
impact on the structure of the Yukawa matrix relevant for the
generation of neutrino masses. The new states in the Zee model are
summarized in Tab.~\ref{tab:Z}. With them, the Yukawa Lagrangian of
the model includes
\begin{equation}
- \mathcal{L}^Y_{\rm Z} = \overline{L} \left( y_e \, H + \Gamma_e \, \phi \right) \, e_R + y_s \, s \, \overline{L^c} \, L + \hc \, ,
\label{Zlagrangian}
\end{equation}
where flavor indices have been omitted. The $3 \times 3$ Yukawa matrix
$y_s$ is antisymmetric in flavor space while $y_e$ and $\Gamma_e$ are
two general $3 \times 3$ complex matrices. In the general THDM, both
Higgs doublets could acquire non-zero VEVs. However, with no quantum
number distinguishing $H$ and $\phi$, one can choose to go to the
so-called \textit{Higgs basis}, in which only one of the two fields 
acquires a VEV. We choose that the electroweak VEV $v$ is obtained as
$v^2 = v_H^2$. In this basis, the expressions for the mass matrices
become especially simple. In case of the charged leptons, this reads
\begin{equation}
\mathcal{M}_e = \frac{v}{\sqrt{2}} \, y_e \, . 
\label{eq:emassZ}
\end{equation}
In the following, and without loss of generality, we will work in the
basis in which $\mathcal{M}_e$ is diagonal. The scalar potential of
the Zee model includes the trilinear term
\begin{equation}
\mathcal V_{\rm Z} \supset \mu_{\rm Z} \, H \, \phi \, s^\ast + \hc \, ,
\label{Zpotential}
\end{equation}
where $\mu_{\rm Z}$ is a parameter with dimensions of mass. After
electroweak symmetry breaking, this trilinear coupling leads to mixing
between the usual charged Higgs of the THDM and $s \equiv s^+$. The
mixing angle, denoted as $\varphi$, is given by
\begin{equation}
s_{2 \varphi} = \sin 2 \varphi = \frac{\sqrt{2} \, v \, \mu_{\rm Z}}{m_{h_2^+}^2 - m_{h_1^+}^2} \, ,
\end{equation}
where $m_{h_1^+}^2$ and $m_{h_2^+}^2$ are the squared masses of the
two physical charged scalars in the spectrum, $h_1^+$ and $h_2^+$,
respectively. The relevance of the trilinear $\mu_{\rm Z}$ goes beyond
this mixing in the charged scalar sector. It is straightforward to
show that a conserved lepton number cannot be defined in the presence
of the Lagrangian terms in Eqs.~\eqref{Zlagrangian} and
\eqref{Zpotential}. In fact, lepton number is explicitly violated in
two units, leading to the generation of Majorana neutrino masses at
the 1-loop level, as shown in Fig.~\ref{fig:Z}. The neutrino mass
matrix is calculable and given by
\begin{equation}
  m_{\rm Z} = - \frac{s_{2 \varphi}}{16 \, \pi^2} \,
    \left( y_s \, \mathcal{M}_e \, \Gamma_e + \Gamma_e^T \, \mathcal{M}_e \, y_s^T \right) \, \log \left( \frac{m_{h_2^+}^2}{m_{h_1^+}^2}\right) \, .
\label{eq:numassZ}
\end{equation}
Direct comparison with the master formula in Eq.~\eqref{eq:master}
indicates that in the Zee model one has $y_1 \ne y_2$. In fact, the
Zee model constitutes a well-known example of a model in which one of
the Yukawa matrices is antisymmetric while the other is a general
complex matrix.

\subsection{The Zee-Babu model}
\label{subsec:zeebabu}

{
\renewcommand{\arraystretch}{1.4}
\begin{table}
\centering
{\setlength{\tabcolsep}{0.5em}
\begin{tabular}{| c c c c c c |}
\hline  
 & spin & generations & $\mathrm{SU(3)}_c$ & $\mathrm{SU(2)}_L$ & $\mathrm{U(1)}_Y$ \\
\hline
\hline    
$s$ & 0 & 1 & ${\bf 1}$ & ${\bf 1}$ & $1$ \\ 
$k$ & 0 & 1 & ${\bf 1}$ & ${\bf 1}$ & $2$ \\ 
\hline
\end{tabular}
}
\caption{New particles in the Zee-Babu model.}
\label{tab:ZB}
\end{table}
}

\begin{figure}[t]
\centering
\includegraphics[width=0.55\textwidth]{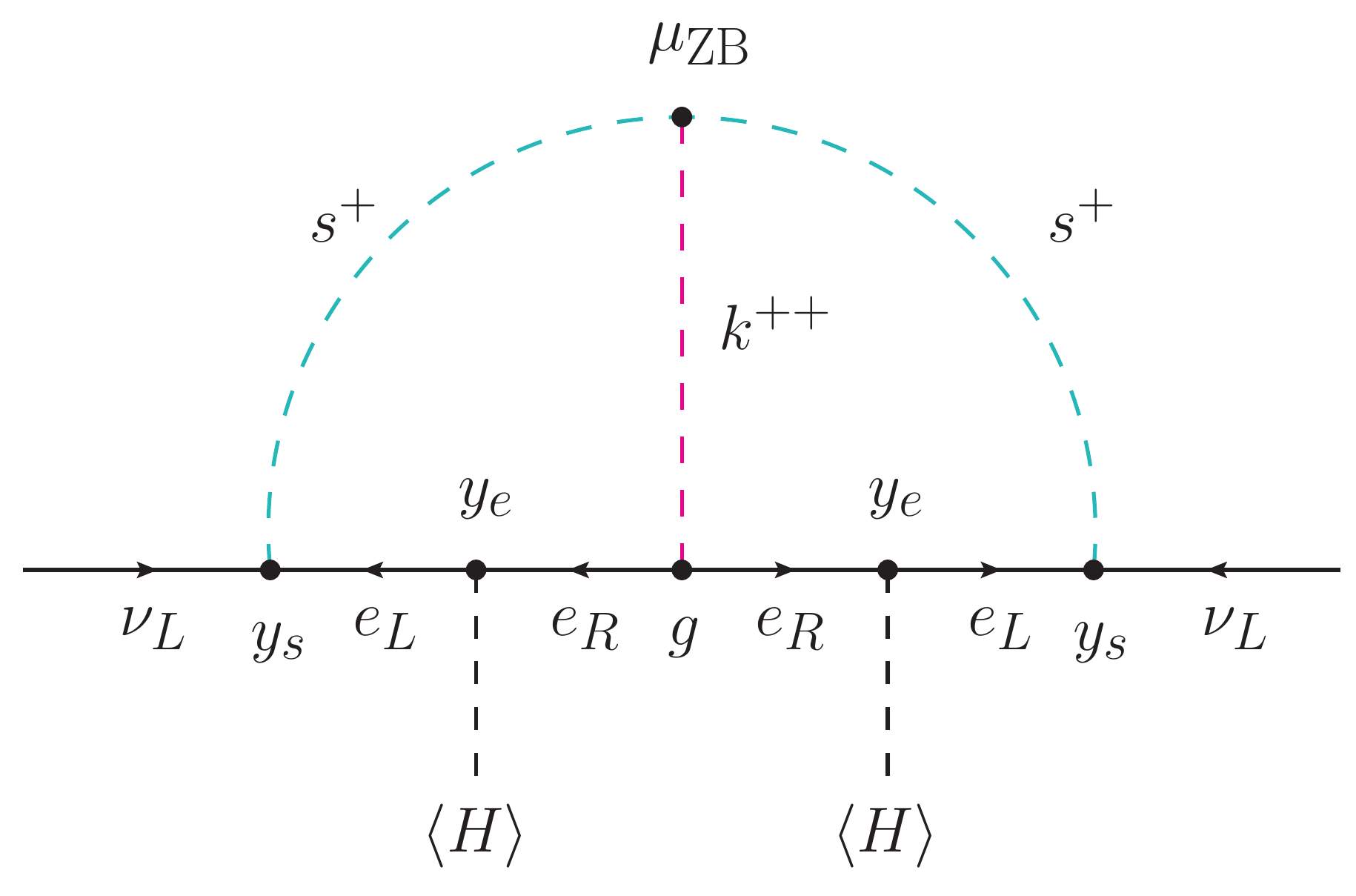}
\caption{Neutrino mass generation in the Zee-Babu model.}
\label{fig:ZB}
\end{figure}

The Zee-Babu model \cite{Cheng:1980qt,Zee:1985id,Babu:1988ki} is a
simple extension of the scalar content of the SM. In addition to the
usual Higgs doublet, two $\rm SU(2)_L$ singlet scalars are introduced:
the singly-charged $s \equiv s^+$ and the doubly-charged $k \equiv
k^{++}$. This is explicitly summarized in Tab.~\ref{tab:ZB}. With
these fields, the Lagrangian includes two new Yukawa terms
\begin{equation}
- \mathcal{L}^Y_{\rm ZB} = y_s \, s \, \overline{L^c} \, L + g \, k \, \overline{e_R^c} \, e_R + \hc \, ,
\label{ZBlagrangian}
\end{equation}
where flavor indices have been omitted. Here $y_s$ is an antisymmetric
$3 \times 3$ Yukawa matrix while $g$ is a symmetric $3 \times 3$
matrix. In addition, the scalar potential of the model includes the
trilinear term
\begin{equation}
\mathcal V_{\rm ZB} \supset \mu_{\rm ZB} \, s \, s \, k^\ast + \hc \, ,
\label{ZBpotential}
\end{equation}
where $\mu_{\rm ZB}$ is a parameter with dimensions of mass. The
simultaneous presence of the Lagrangian terms in
Eqs.~\eqref{ZBlagrangian} and \eqref{ZBpotential} implies the breaking
of lepton number in two units. This leads to the generation of
Majorana neutrino masses at the 2-loop level, as shown in
Fig.~\ref{fig:ZB}. In this graph $y_e$ is the SM lepton Yukawa term,
defined as $y_e \, H \, \overline{L} \, e_R$. The resulting expression
for the neutrino mass matrix takes the form
\begin{equation}
m_{\rm ZB}= \frac{v^2 \, \mu_{\rm ZB}}{(16 \pi^2)^2 \, m_s^2} \, y_s \, y_e \, g \, y_e^T \, y_s^T \, F_{\rm ZB}\left( \frac{m_k^2}{m_s^2}\right) \, ,
\label{eq:numassZB}
\end{equation}
where $m_s$ and $m_k$ are the $s$ and $k$ squared masses,
respectively, and $F_{\rm ZB}$ is a dimensionless loop
function. Therefore, we see that in the Zee-Babu model one has
$\sqrt{2} \, y_1 = \sqrt{2} \, y_2 = y_s$, with $y_s = - y_s^T$. This indeed
implies a prediction: since $\text{Det}(y_s) = 0$, one of the neutrinos
remains massless.

\section{Summary}
\label{sec:summary}

We have presented a general parametrization for the Yukawa couplings
in Majorana neutrino mass models. We call this the master
parametrization.  A proof for the master parametrization has also been
presented, see Appendix~\ref{sec:proof}. In order to help the reader
in practical applications, we have also provided a {\tt Mathematica}
notebook that implements the master parametrization
in~\cite{masterweb}. The aim of this master parametrization is to
generalize the well-known Casas-Ibarra parametrization, which in its
strict original form is valid only for the type-I seesaw. Although
different adaptations of the Casas-Ibarra parametrization have been
discussed in the context of concrete models in the literature, the aim
of our master parametrization is to be as completely general as
possible.

We stress that our master parametrization is valid for any Majorana
neutrino mass model. We have shown its application to various
well-known example models. We have also discussed some particular
cases, where the Yukawa couplings are no longer completely free
parameters but, typically for symmetry reasons, have to obey some
restrictions. In such cases, the application of the master parametrization
may become either trivial or impractically complicated, depending
on the complexity of the extra conditions, as we discussed with
some examples.

Let us briefly mention that from the list of examples that we
have discussed in Section~\ref{sec:apps}, one should not derive the
\textit{incorrect} conclusion that only very few neutrino mass models
require the full power of the master parametrization.  This bias in
our example list is mainly due to the fact that in our discussion
we have focused on the best-known neutrino mass models that exist in
the literature.

\begin{figure}[t]
\centering
\includegraphics[width=0.5\textwidth]{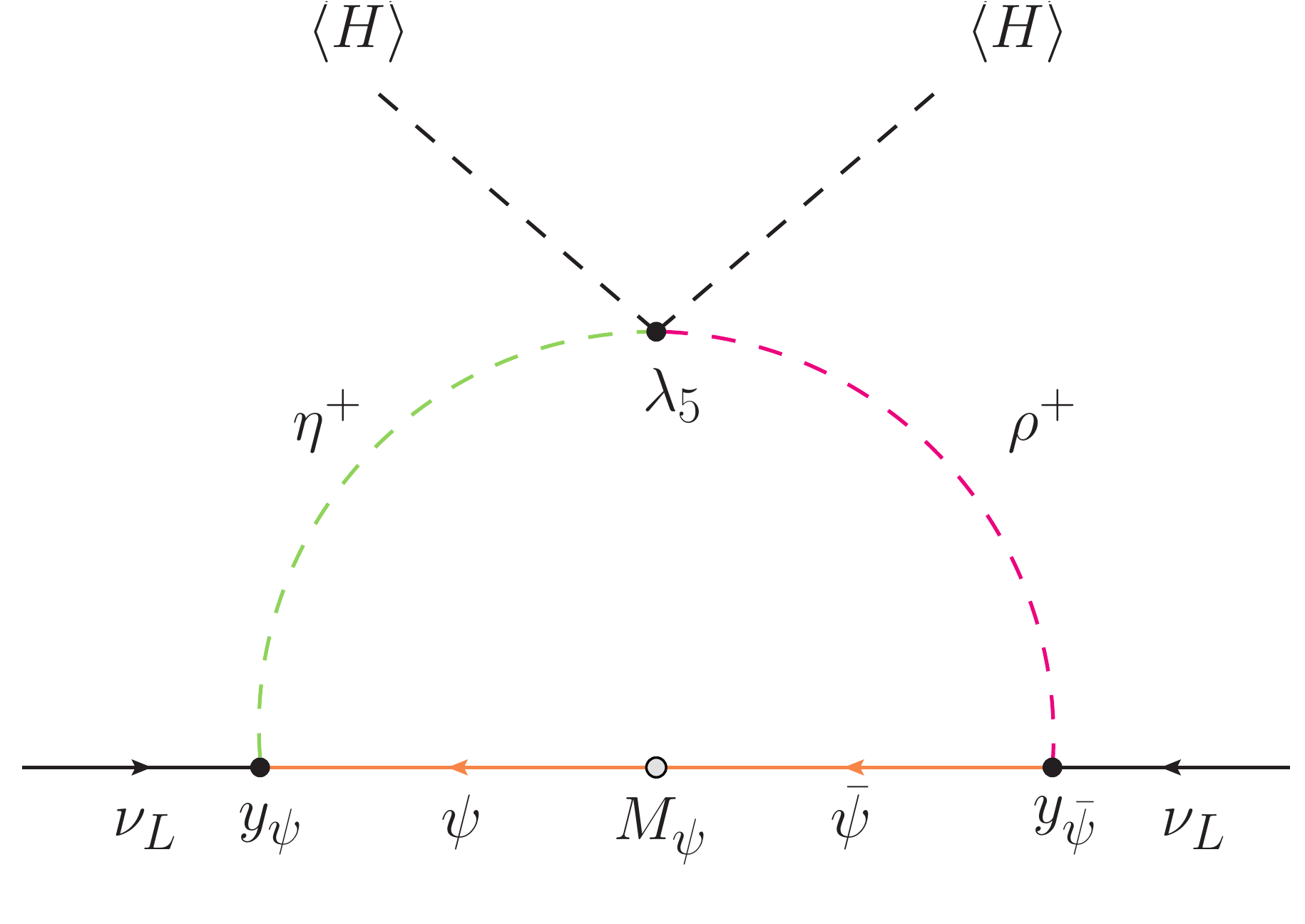}
\caption{Neutrino mass generation in a modified scotogenic model, for
  defintions see text. This simple variation necessarily has two
  independent Yukawa couplings.}
\label{fig:SCMod}
\end{figure}

In fact, once one goes beyond the minimal $d=5$ tree-level
realizations of the Weinberg operator, the majority of models have
$y_1 \ne y_2$ and the Casas-Ibarra parametrization can not cover these
models, as we have stressed several times.  At tree-level, at $d=7$ we
find the BNT model~\cite{Babu:2009aq} at $d=9$ one of the two genuine
models (model-II) in~\cite{Anamiati:2018cuq} is also of this type.
Actually, for radiative neutrino mass models the majority of models
are of this type. This can be easily understood as follows.  Consider,
for example, the neutrino mass model shown in Fig.~\ref{fig:SCMod}.
The diagram is the same as in the scotogenic model. Here, the
vector-like fermion $\psi$ transforming as $({\bf 1},{\bf 3},1)$ (with
its vector partner ${\bar\psi}$) replace the singlet fermions of the
original model. In addition to $\eta$, a second doublet $\rho$ with
quantum numbers $({\bf 1},{\bf 2},3/2)$ is introduced. This model
obviously has two independent Yukawa couplings and thus, the full
master parametrization is needed to describe its parameter space.
Another example of a modified Scotogenic model with $y_1 \ne y_2$ can
be found in~\cite{Ma:2013yga,Hagedorn:2018spx}. Unsurprisingly, at
loop level there are actually more variations with this type of
``asymmetric'' diagrams, i.e. $y_1 \ne y_2$, than variations with
``symmetric'' diagrams (where the field coupling to the two neutrinos
is necessarily the same) as can be seen, for example, in the tables
of~\cite{Bonnet:2012kz,Sierra:2014rxa} or the list of diagrams at d=7
1-loop in~\cite{Cepedello:2017eqf}.

We close by mentioning again that we have concentrated our discussion
on the particular case of three, light active neutrinos. It is
possible to extend our approach to four or more neutrinos, if ever
this becomes necessary. Technically, the form of our master
parametrization would remain the same, but the dimensions of the
defining matrices will change, and the explicit forms of the matrices
$C_1$ and $C_2$, defined in Section~\ref{sec:master}, would need to be
calculated.

\section*{Acknowledgements}

Work supported by the Spanish grants PGC2018-095984-B-I00 and
FPA2017-85216-P (AEI/FEDER, UE), PROMETEO/2018/165, PROMETEO/2019/071
and SEJI/2018/033 (Generalitat Valenciana) and the Spanish Red
Consolider MultiDark FPA2017-90566-REDC.

\appendix

\section{Proof}
\label{sec:proof}

In the following, we provide a constructive proof of the master
parametrization. We begin by replacing the Takagi decomposition of $m$
in Eq.~\eqref{eq:m-U} and the singular-value decomposition of $M$ in
Eq.~\eqref{eq:M-SVD} into the master formula in
Eq.~\eqref{eq:master}. Moreover, we take $f=1$ to simplify the
expressions in the proof, and take this global factor into account by
rescaling the final expressions for the $y_1$ and $y_2$ Yukawa
matrices. This leads to
\begin{equation}
(U^\dagger)^T \, D_m \, U^\dagger = y_1^T \, V_1^T \, \widehat \Sigma \, V_2 \, y_2 
+ y_2^T \, V_2^T \, \widehat \Sigma^T \, V_1 \, y_1 \, .
\end{equation}
Multiplying the previous expression on the left by $(\bar{D}_{\sqrt{m}})^{-1} \, U^T$ 
and on the right by $U \, (\bar{D}_{\sqrt{m}})^{-1}$, with $\bar{D}_{\sqrt{m}}$ 
introduced in Eq.~\eqref{eq:defDsqrtm}, one obtains
\begin{gather}  
(\bar{D}_{\sqrt{m}})^{-1} \, D_{\sqrt{m}} \, D_{\sqrt{m}} \, (\bar{D}_{\sqrt{m}})^{-1} 
= \nonumber \\
(\bar{D}_{\sqrt{m}})^{-1} \, U^T y_1^T \, V_1^T \, \widehat \Sigma \, V_2 \, y_2 \, 
U \, (\bar{D}_{\sqrt{m}})^{-1} + (\bar{D}_{\sqrt{m}})^{-1} \, U^T y_2^T \, V_2^T \, 
\widehat \Sigma^T \, V_1 \, y_1 \, U \, (\bar{D}_{\sqrt{m}})^{-1} \, .
\label{eq:intermediate1}
\end{gather}
This expression clearly suggests to introduce
\begin{align}
\bar{y}_1 &= V_1 \, y_1 \, U \, (\bar{D}_{\sqrt{m}})^{-1}\, , \label{eq:ybar1} \\
\bar{y}_2 &= V_2 \, y_2 \, U \, (\bar{D}_{\sqrt{m}})^{-1}\, . \label{eq:ybar2}
\end{align}
We note that $y_1$ and $y_2$ can be univocally determined from
$\bar{y}_1$ and $\bar{y}_2$, since all the other matrices
participating in Eqs.~\eqref{eq:ybar1} and \eqref{eq:ybar2} are
invertible. With these definitions, Eq.~\eqref{eq:intermediate1} is
equivalent to
\begin{equation}
\bar{y}_1^T \, \widehat \Sigma \, \bar{y}_2 + \bar{y}_2^T \, \widehat \Sigma^T \, \bar{y}_1 
= \left\{ \begin{array}{cr}\id_3 \quad & \text{if} \ r_m = 3 \, , \\ & \\ 
P \, \begin{pmatrix} 0 & 0 & 0 \\ 0 & 1 & 0 \\ 0 & 0 & 1 \end{pmatrix} \, P
\quad & \text{if} \ r_m = 2 \, . 
\end{array} \right.
\label{eq:intermediate2}
\end{equation}
In the next step, we write the matrices $\bar{y}_1$, $\bar{y}_2$ and
$\widehat \Sigma$ in blocks. As we will see below, this will allow us
to identify some arbitrary blocks and focus the discussion on the
non-trivial ones. Using the general expression for $\widehat \Sigma$
given in Eq.~\eqref{eq:sigma}, the combination $\bar{y}_1^T \,
\widehat \Sigma \, \bar{y}_2 + \bar{y}_2^T \, \widehat \Sigma^T \,
\bar{y}_1$ in Eq.~\eqref{eq:intermediate2} can be written as
\begin{gather}
\bar{y}_1^T \, \widehat \Sigma \, \bar{y}_2 + \bar{y}_2^T \, \widehat \Sigma^T \, \bar{y}_1 = \nonumber \\
\left((\bar{y}_1)_n^T \;\; (\bar{y}_1)_{n_2-n}^T \;\; (\bar{y}_1)_{n_1-n_2}^T\right)
\left(\begin{array}{c} \begin{array}{cc} \Sigma & 0 \\ 0 & 0_{n_2-n} \end{array} \\ \hline 0_{n_1-n_2} \end{array}\right)
\left(\begin{array}{c}(\bar{y}_2)_n \\ (\bar{y}_2)_{n_2-n} \end{array}\right) + \left((\bar{y}_2)_n^T \;\; (\bar{y}_2)_{n_2-n}^T\right)
\left(\begin{array}{c|c} \begin{array}{cc} \Sigma & 0 \\ 0 & 0^T_{n_2-n} \end{array} & 0^T_{n_1-n_2} \end{array}\right)
\left(\begin{array}{c}(\bar{y}_1)_n \\ (\bar{y}_1)_{n_2-n} \\ (\bar{y}_1)_{n_1-n_2} \end{array}\right) \nonumber \\
= (\bar{y}_1)_n^T \, \Sigma \, (\bar{y}_2)_n + (\bar{y}_2)_n^T \, \Sigma \, (\bar{y}_1)_n \, .
\end{gather}
We clearly see that there are some blocks which can have arbitrary
values since they multiply null matrices and drop in the final
expression. These are
\begin{align}
X_1 &= \sqrt{2} \, (\bar{y}_1)_{n_2-n}\in\mathbb{C}^{(n_2-n)\times 3} \, , \nonumber \\
X_2 &= \sqrt{2} \, (\bar{y}_1)_{n_1-n_2}\in\mathbb{C}^{(n_1-n_2)\times 3} \, , \nonumber \\
X_3 &= \sqrt{2} \, (\bar{y}_2)_{n_2-n}\in\mathbb{C}^{(n_2-n)\times 3} \, ,
\end{align}
where the $\sqrt{2}$ factors have been introduced for
convenience. $X_1$, $X_2$ and $X_3$ have $6(n_2-n)$, $6(n_1-n_2)$ and
$6(n_2-n)$ free real parameters, respectively. We define now
\begin{align}
\bar{\bar{y}}_1 &= \sqrt{2}\,\Sigma^{1/2} \, (\bar{y}_1)_n \, , \\
\bar{\bar{y}}_2 &= \sqrt{2}\,\Sigma^{1/2} \, (\bar{y}_2)_n \, .
\end{align}
Again, $\bar{y}_1$ and $\bar{y}_2$, and hence the original Yukawa
matrices $y_1$ and $y_2$, can be univocally obtained from
$\bar{\bar{y}}_1$ and $\bar{\bar{y}}_2$ since the matrix
$\Sigma^{1/2}$ is invertible. With these redefinitions,
Eq.~\eqref{eq:intermediate2} is equivalent to
\begin{equation}
\bar{\bar{y}}_1^T \, \bar{\bar{y}}_2 + \bar{\bar{y}}_2^T \, \bar{\bar{y}}_1 =
\left\{ \begin{array}{cr} 2 \, \id_3 \quad & \text{if} \ r_m = 3 \, , \\ & \\ 
2 \, P \, \begin{pmatrix} 0 & 0 & 0 \\ 0 & 1 & 0 \\ 0 & 0 & 1 \end{pmatrix}  \, P 
\quad & \text{if} \ r_m = 2 \, . 
\end{array} \right.
\label{eq:intermediate3}
\end{equation}
At this point, the roles of $\bar{\bar{y}}_1$ and $\bar{\bar{y}}_2$
are completely interchangeable. Therefore, without loss of generality,
we will first determine the form of $\bar{\bar{y}}_1$ and then derive
$\bar{\bar{y}}_2$. We define
$r=\mathrm{rank}(\bar{\bar{y}}_1)$. Equivalently, the matrix
$\bar{\bar{y}}_1$ contains $r$ linearly independent columns. It
follows
\begin{equation} \label{eq:r}
1 \leq r \leq \min(n,3) \leq 3 \, ,
\end{equation}
simply because $\bar{\bar{y}}_1$ is a non-null $n \times 3$
matrix. $\bar{\bar{y}}_1$ can now be written as the product of an
$n\times r$ matrix, with $r$ orthogonal columns, and a matrix with
vanishing entries below the main diagonal. That is, there exists a
matrix $W \in \mathbb{C}^{n\times r}$, with $\text{rank}(W)=r$ and
$W^\dagger W = W^T W^* = \id_r$, and a matrix $A\in\mathbb{C}^{r\times
  3}$, such that
\begin{equation} \label{eq:QR}
\bar{\bar{y}}_1 = W \, A \, .
\end{equation}
For the particular case $r = 3$, $A$ is a square upper triangular
matrix, but in general $A$ is a rectangular matrix with vanishing
entries below the main diagonal. The factorization in
Eq.~\eqref{eq:QR} is unique provided some conditions on $A$ are
satisfied. These conditions depend on the values of $r$ and $r_m$ and
will be discussed below for each case. The matrix $W$, whose $r$
columns are orthogonal, can be completed to form an orthonormal basis
of $\mathbb{C}^{n\times n}$, resulting in the $n \times n$ unitary
matrix $\widehat W$, given by
\begin{equation}
\widehat W = (W \; \bar{W}) \, .
\end{equation}
Although the completion of the basis (and thus the matrix
$\bar{W}\in\mathbb{C}^{n\times(n-r)}$) is not uniquely defined, the
vector subspace that it spans is, and this suffices for the rest of
this proof. We now derive the implications for the matrix
$\bar{\bar{y}}_2$ given this form for $\bar{\bar{y}}_1$. The matrix
$\bar{\bar{y}}_2$ can be written in terms of the basis $\widehat W^*$ 
as
\begin{equation}
\bar{\bar{y}}_2 = \widehat W^*\,\widehat B = (W^* \; \bar{W}^*) \left(\begin{array}{c} B \\ \bar{B} \end{array}\right) \, ,
\end{equation}
with $\bar{B}\in\mathbb{C}^{(n-r)\times 3}$ an arbitrary matrix
containing $6(n-r)$ real free parameters. We note that this matrix is
indeed completely arbitrary due to the fact that it drops in the
products $\bar{\bar{y}}_1^T \, \bar{\bar{y}}_2$ and $\bar{\bar{y}}_2^T
\, \bar{\bar{y}}_1$ since $\bar{W^\dagger} \, W = 0$. With this
definition, Eq.~\eqref{eq:intermediate3} becomes
\begin{equation} \label{eq:AB}
A^T \, B + B^T \, A = 
\left\{ \begin{array}{cc} 2 \, \id_3
\quad & \text{if} \ r_m = 3 \, , \\ & \\
2 \, P \, \begin{pmatrix} 0 & 0 & 0 \\ 0 & 1 & 0 \\ 0 & 0 & 1 \end{pmatrix} \, P
\quad & \text{if} \ r_m = 2 \, . 
\end{array} \right.
\end{equation}
This constraint on the $r\times 3$ matrices $A$ and $B$ is completely
equivalent to the master formula in Eq.~\eqref{eq:master}. Therefore,
we just need to determine $A$ and $B$ and the master parametrization
will be finally obtained. In the following, the proof for the case
$r_m=2$ will assume NH, and thus $P=\id_3$. The IH case, with
$P=P_{13}$, will be recovered a posteriori with the substitutions $A
\to A \, P_{13}$ and $B \to B \, P_{13}$. In order to find $A$ and $B$
it proves convenient to express them in terms of some auxiliary
matrices, to be determined by imposing Eq.~\eqref{eq:AB}. First, $A$
can be written as
\begin{equation} \label{eq:Afact}
A = T \, C_1,
\end{equation}
where $T \in\mathbb{C}^{r \times r}$ is a general upper-triangular
invertible square matrix with positive real diagonal entries and $C_1
\in\mathbb{C}^{r\times 3}$ is a matrix that must be
determined.~\footnote{We can recover the IH scenario, with
  $P=P_{13}$, by replacing $A \to A \, P_{13}$ and $B
  \to B \, P_{13}$ or, equivalently, $C_1 \to C_1 \, P_{13}$.}  This
factorization of the matrix $A$ is always possible and singles out the
upper triangular square matrix $T$.  Regarding $B$, it can be
expressed as
\begin{equation} \label{eq:Bexpression}
B \equiv B\left( T , K , C_1 , C_2 \right) = \left( T^T \right)^{-1} 
\, \left[ C_1 \, C_2 + K \, C_1 \right] \, ,
\end{equation}
where $K \in\mathbb{C}^{r\times r}$ is an antisymmetric $r \times r$
matrix and $C_2\in\mathbb{C}^{3\times 3}$ must be
determined.~\footnote{Again, we point out that for $r_m=2$ we focus on
  NH with $P=\id_3$. The IH scenario is obtained by making the
  replacements $A \to A \, P_{13}$ and $B \to B \, P_{13}$, equivalent
  to $C_2 \to P_{13}\,C_2\,P_{13}$.} This form for the matrix $B$ can
be justified by direct computation. One always finds that the
resulting $B$ matrix can be written in this way, with the specific
forms for the $C_1$ and $C_2$ matrices depending on $r$ and $r_m$. In
fact, the rest of the proof consists in obtaining specific expressions
for $C_1$ and $C_2$ compatible with Eq.~\eqref{eq:AB}. In order to
cover all scenarios, we will consider all possible $r$ and $r_m$
values, and denote them with the pair of numbers $(r_m,r)$. Let us now
explore all the different possibilities one by one.

\begin{center}
$\boxed{\boldsymbol{r=3}}$
\end{center}

In this case, $\bar{\bar{y}}_1$ contains $3$ linearly independent
columns and $A\in\mathbb{C}^{3\times 3}$ is an upper triangular
invertible square matrix. One can simply write $A$ as
\begin{equation} \label{eq:Acase1}
A = \begin{pmatrix}\alpha_{11} & \alpha_{12} & \alpha_{13} \\
0 & \alpha_{22} & \alpha_{23} \\ 0 & 0 & \alpha_{33} \end{pmatrix} \, ,
\end{equation}
with $\alpha_{11}$, $\alpha_{22}$ and $\alpha_{33}$ real positive
values. Since $A$ is a square matrix, one can identify $A = T$. One
can now distinguish two sub-cases depending on the value of $r_m$.

\begin{itemize}
\item[$\ast$] \textit{Case $(3,3)$: $r_m = 3$}
\end{itemize}

For $r_m = 3$, the identification $A = T$ allows one to conclude that
\begin{equation}
C_1 = \id_3 \, .
\end{equation}
In fact, in this case Eq.~\eqref{eq:QR} is the QR decomposition of the
matrix $\bar{\bar{y}}_1$. One can now replace the expressions for the
$A$ and $B$ matrices, including the identification $C_1 = \id_3$, into
Eq.~\eqref{eq:AB}. This direct computation leads to
\begin{equation}
C_2 = \id_3 \, .
\end{equation}

\begin{itemize}
\item[$\ast$] \textit{Case $(2,3)$: $r_m = 2$}
\end{itemize}

Alternatively, if $r_m=2$, and taking into account the possible values
of the matrix $P$, it can be easily shown that one gets
\begin{equation}
C_1=P \quad , \quad C_2 = P \, \begin{pmatrix} 0 & 0 & 0 \\ 0 & 1 & 0 \\ 0 & 0 & 1 \end{pmatrix} \, P \, .
\end{equation}

In both sub-cases, the matrices $W$, $T$ and $K$ have, respectively,
$6n-9$, $9$ and $6$ free real parameters.

\begin{center}
$\boxed{\boldsymbol{r=2}}$
\end{center}

In this case we consider three scenarios. They differ in the way the
rank $r$ gets reduced to $2$.

\begin{itemize}
\item {\bf $\boldsymbol{r=2}$, with linearly independent second
  and third columns of $\boldsymbol{\bar{\bar{y}}_1}$}
\end{itemize}

As in case 2, $\bar{\bar{y}}_1$ contains $2$ linearly independent
columns and $A\in\mathbb{C}^{2\times 3}$ is a rectangular matrix with
the form
\begin{equation} \label{eq:Acase3}
A = \begin{pmatrix}\alpha_{11} & \alpha_{12} & \alpha_{13} \\ 
\alpha_{21} & 0 & \alpha_{23} \end{pmatrix} \, ,
\end{equation}
with $\alpha_{12}$ and $\alpha_{23}$ positive real values. We can
write $A = T \,C_1$ and distinguish again two sub-cases depending on
the value of $r_m$.

\begin{itemize}
\item[$\ast$] \textit{Case $(3,2)_a$: $r_m = 3$}
\end{itemize}

Again, we replace the general expressions for the $A$ and $B$
matrices, adapted in this case to $r = 2$ and $r_m = 3$, into
Eq.~\eqref{eq:AB}. One obtains, simply by direct computation, 
that the matrix $C_1$ must have the form 
\begin{equation}
C_1 = \begin{pmatrix} z_1 & 1 & 0 \\ z_2 & 0 & 1 \end{pmatrix} \, ,
\end{equation}
with $z_1$ and $z_2$ two complex numbers such that $1+z_1^2+z_2^2=0$,
while $C_2$ is given by
\begin{equation} \label{C23a}
C_2 = \begin{pmatrix} -1 & 0 & 0 \\ 0 & 1 & 0 \\ 0 & 0 & 1 \end{pmatrix} \, .
\end{equation}

\begin{itemize}
\item[$\ast$] \textit{Case $(2,2)_a$: $r_m = 2$}
\end{itemize}

If $r_m=2$, one finds analogous expressions for the matrices $C_1$ and
$C_2$,
\begin{equation} \label{C13a}
C_1 = \begin{pmatrix} z_1 & 1 & 0 \\ z_2 & 0 & 1 \end{pmatrix} \, P \quad , \quad C_2 = P \, \begin{pmatrix} -1 & 0 & 0 \\ 0 & 1 & 0 \\ 0 & 0 & 1 \end{pmatrix} \, P \, .
\end{equation}
However, in this sub-case it can be shown that $z_1$ and $z_2$ must
obbey the relation $z_1^2+z_2^2=0$. \\

In both sub-cases, the matrix $W$ contains $4(n-1)$ real free
parameters. Moreover, the matrix $T$ has 4 while $K$ has 2. One also
finds two additional real parameters in $C_1$ ($z_1$ or $z_2$).

\begin{itemize}
\item {\bf $\boldsymbol{r=2}$, with a non-null third column of
  $\boldsymbol{\bar{\bar{y}}_1}$, and linearly dependent second and
  third columns of $\boldsymbol{\bar{\bar{y}}_1}$}
\end{itemize}

In this case, $\bar{\bar{y}}_1$ contains $2$ linearly independent
columns and $A\in\mathbb{C}^{2\times 3}$ is a rectangular matrix with
the form
\begin{equation}
A = \begin{pmatrix} \alpha_{11} & \alpha_{12} & \alpha_{13} \\ 
\alpha_{21} & 0 & 0 \end{pmatrix} \, ,
\end{equation}
with $\alpha_{21}$ and $\alpha_{13}$ real positive values. Again, we can write 
$A=T \,C_1$ and particularize the analysis depending on $r_m$.

\begin{itemize}
\item[$\ast$] \textit{Case $(3,2)_b$: $r_m = 3$}
\end{itemize}

If $r_m=3$, one finds by direct computation
\begin{equation}
C_1=\begin{pmatrix} 0 & \pm i & 1 \\ 1 & 0 & 0 \end{pmatrix} \quad , \quad
C_2=\begin{pmatrix} 1 & 0 & 0 \\ 0 & -1 & 0 \\ 0 & 0 & 1 \end{pmatrix} \, .
\end{equation}

\begin{itemize}
\item[$\ast$] \textit{Case $(2,2)_b$: $r_m = 2$}
\end{itemize}

One obtains analogous expressions as for $r_m = 3$,
\begin{equation}
C_1=\begin{pmatrix} 0 & \pm i & 1 \\ 1 & 0 & 0 \end{pmatrix} \, P \quad , \quad
C_2 = P \, \begin{pmatrix} 0 & 0 & 0 \\ 0 & -1 & 0 \\ 0 & 0 & 1 \end{pmatrix} \, P \, .
\end{equation}

In both sub-cases, the matrix $W$ contains $4(n-1)$ free real
parameters, $T$ $4$ and $K$ $2$.

\begin{itemize}
\item {\bf $\boldsymbol{r=2}$, with a third column of $\boldsymbol{\bar{\bar{y}}_1}$ full of zeros}
\end{itemize}

In this case, $\bar{\bar{y}}_1$ contains $2$ linearly independent
columns and $A\in\mathbb{C}^{2\times 3}$ is a rectangular matrix with
the form
\begin{equation} \label{eq:Acase2}
A = \begin{pmatrix} \alpha_{11} & \alpha_{12} & 0 \\ 0 & \alpha_{22} & 0 \end{pmatrix} \, ,
\end{equation}
with $\alpha_{11}$ and $\alpha_{22}$ real positive values. In
principle, we could replace this form for $A$ into Eq.~\eqref{eq:AB},
find that $B$ can be written as in Eq.~\eqref{eq:Bexpression} and
determine $C_1$ and $C_2$. However, it is easy to see that this case
is not compatible with Eq.~\eqref{eq:AB}. If all the entries of the
third column of $\bar{\bar{y}}_1$ vanish, $\left(A^T B + B^T
A\right)_{33} = 0$, and this is clearly not compatible with
Eq.~\eqref{eq:AB}, which requires that element to be $2$ in case of
$r_m = 3$. One also reaches a contradition in case of $r_m = 2$. The
NH case is completely equivalent, whereas the IH case, obtained with
the replacements $A \to A \, P$, $B \to B \, P$, leads to $\left(A^T B
+ B^T A\right)_{11} = 0$, again in contradiction with
Eq.~\eqref{eq:AB}.

\newpage

\begin{center}
$\boxed{\boldsymbol{r=1}}$
\end{center}

In this case, $\bar{\bar{y}}_1$ contains only $1$ linearly independent
column and $A\in\mathbb{C}^{1\times 3}$ is a $1 \times 3$ rectangular
matrix, or equivalently a row vector, with the form
\begin{equation}
A = \begin{pmatrix} \alpha_{11} & \alpha_{12} & \alpha_{13} \end{pmatrix} \, .
\end{equation} 
We now particularize for $r_m$.

\begin{itemize}
\item[$\ast$] \textit{Case $(3,1)$: $r_m = 3$}
\end{itemize}

For $r_m=3$ one can first inspect the diagonal elements of the
equation $A^T B + B^T A = 2 \, \id_3$ and get
\begin{equation} \label{eq:alphabeta}
2 \, \alpha_{1i} \, \beta_{1i} = 2 \, ,
\end{equation}
where the elements of the $B$ matrix are denoted by
$\beta_{ij}$. Eq.~\eqref{eq:alphabeta} is equivalent to $\alpha_{1i}
\neq 0 \neq \beta_{1i}$ and
\begin{equation}
\beta_{1i} = \frac{1}{\alpha_{1i}} \, .
\end{equation}
However, one can now inspect the non-diagonal elements of the equation
$A^T B + B^T A = 2 \, \id_3$. In this case one gets the relations
\begin{equation}
\alpha_{12}^2 = -\alpha_{11}^2 = \alpha_{13}^2 = -\alpha_{12}^2 \, ,
\end{equation}
which imply $\alpha_{12}=0$. Since this contradicts our previous
deduction we conclude that there is no possible solution in this
sub-case: $r_m = 3$ is not compatible with $r = 1$.

\begin{itemize}
\item[$\ast$] \textit{Case $(2,1)$: $r_m = 2$}
\end{itemize}

One can see that for $r_m=2$ there is a redundant equation (or $2$ 
redundant real equations). One also finds that Eq.~\eqref{eq:AB} 
leads to $T=\alpha_{13}\neq 0$ ($T=\alpha_{11}\neq 0$ in the IH case with
$P=P_{13}$), that can be considered a positive real value, $A = T \, C_1$, with
\begin{equation}
C_1 = \begin{pmatrix} 0 & \pm i & 1 \end{pmatrix} \, P \quad , \quad
C_2 = P \, \begin{pmatrix} 0 & 0 & 0 \\ 0 & -1 & 0 \\ 0 & 0 & 1 \end{pmatrix} \, P.
\end{equation}
Moreover, since $r=1$, $K = 0$ vanishes. Due to the latter, the $B$ matrix
receives a simplified form,
\begin{equation}
B \equiv B\left( T , C_1 , C_2 \right) = \frac{1}{T} \, C_1 \, C_2 \, .
\end{equation}
In this subcase $W$ contains $2n-1$ free real parameters and $T$ has 1. \\

This concludes the proof of the master parametrization.

\section{Parametrization of the matrices in the master parametrization}
\label{sec:mat}

Some of the matrices involved in the master parametrization can be
further parametrized in terms of certain real parameters, in some
cases with a clear physical meaning. In this Appendix we collect these
parametrizations, which may be useful in practical applications of the
master parametrization. \\

First, the unitary matrix $U$ is generally parametrized in terms of
three mixing angles and three phases (in case of Majorana neutrinos)
as
\begin{equation}
\label{eq:PMNS}
U=
\left(
\begin{array}{ccc}
 c_{12}c_{13} & s_{12}c_{13} & s_{13}e^{i\delta}  \\
-s_{12}c_{23}-c_{12}s_{23}s_{13}e^{-i\delta} & 
c_{12}c_{23}-s_{12}s_{23}s_{13}e^{-i\delta} & s_{23}c_{13}  \\
s_{12}s_{23}-c_{12}c_{23}s_{13}e^{-i\delta} & 
-c_{12}s_{23}-s_{12}c_{23}s_{13}e^{-i\delta} & c_{23}c_{13}  
\end{array}
\right) \ 
\left(
\begin{array}{ccc}
1 & 0 & 0 \\
0 & e^{i \, \eta_2} & 0 \\
0 & 0 & e^{i \, \eta_3}
\end{array}
\right) \, .
\end{equation}
Here $c_{ij} \equiv \cos \theta_{ij}$ and $s_{ij} \equiv \sin
\theta_{ij}$. The parameter $\delta$ is usually referred to as the
Dirac phase, while $\eta_2$ and $\eta_3$ are the Majorana phases,
since they are only physical in case of Majorana neutrinos. The angles
$\theta_{ij}$ can be taken in the first quadrant, $\theta_{ij} \in
\left[ 0 , \pi/2 \right]$ while the phases $\delta$ and $\eta_{2,3}$
can take any value in the range $\left[ 0 , 2 \, \pi
  \right]$. Furthermore, the three neutrino mass eigenvalues contained
in the matrix $D_m$ can be written in terms of lightest neutrino mass, $m_0$,
and two squared mass differences. In case of neutrino NH they are
given by
\begin{align}
m_1 &= m_0 \, , \\
m_2 &= \sqrt{\Delta m^2_{21} + m_0^2} \, , \\
m_3 &= \sqrt{\Delta m^2_{31} + m_0^2} \, , \\
\end{align}
whereas in case of neutrino IH they follow
\begin{align}
m_1 &= \sqrt{|\Delta m^2_{31}| + m_0^2} \, , \\
m_2 &= \sqrt{|\Delta m^2_{31}| + \Delta m^2_{21} + m_0^2} \, , \\
m_3 &= m_0 \, . \\
\end{align}
Neutrino oscillation experiments are sensitive to the three leptonic
mixing angles, the two squared mass differences and the Dirac
phase. We refer to \cite{deSalas:2017kay} for a state-of-the-art
global fit to these parameters.

The complex unitary $n \times n$ matrix $\widehat W$ can also be
conveniently parametrized. Here we make use of
\cite{hedemann2013hyperspherical}, which discusses the canonical form
for a generic $n \times n$ unitary matrix. In case of the
common case of $n = 3$, $\widehat W$ can be expressed as
\begin{equation}
\label{eq:Wmat}
\widehat W =
\left(
\begin{array}{ccc}
1 & 0 & 0 \\
0 & a & b \\
0 & -b^* & a^* 
\end{array}
\right) \ 
\left(
\begin{array}{ccc}
c & 0 & d \\
0 & 1 & 0 \\
-d^* & 0 & c^*
\end{array}
\right) \ 
\left(
\begin{array}{ccc}
e & f & 0 \\
-f^* & e^* & 0 \\
0 & 0 & 1
\end{array}
\right) \, ,
\end{equation}
with $|a|^2 + |b|^2 = 1$, $|c|^2 + |d|^2 = 1$ and $|e|^2 + |f|^2 =
1$. One has 6 complex parameters but they must satisfy 3 real
conditions. This makes 9 real free parameters, as expected for a $3
\times 3$ unitary matrix.~\footnote{A $n \times n$ unitary matrix
  contains $n^2$ independent real parameters.} Examples for other
values of $n$ can be found in \cite{hedemann2013hyperspherical}.

Finally, in the particular case of $\sqrt{2} \, y_1 = \sqrt{2} \, y_2
= y$ and $n_1 = n_2 = n = r_m = r = 3$, the master formula reduces to
the usual type-I seesaw form $m = y^T M y$ and the master
parametrization to the well-known Casas-Ibarra parametrization. This
allows to write the Yukawa matrix $y$ in terms of low-energy and model
parameters and the so-called Casas-Ibarra $R$ matrix, an orthogonal $3
\times 3$ matrix such that $R^T R = R R^T = \id_3$. This matrix can be
generally parametrized as
\begin{equation} \label{eq:defR}
R  = S \, R_3 \, R_2 \, R_1 \, ,
\end{equation}
with
\begin{equation}\label{eq:defR2}
R_3 =
\begin{pmatrix}
\cos(z_3) & -\sin(z_3) & 0 \\
\sin(z_3) & \cos(z_3) & 0 \\
0 & 0 & 1 
\end{pmatrix} , \quad
R_2 =
\begin{pmatrix}
\cos(z_2) & 0 & -\sin(z_2)  \\
\sin(z_2) & 0 & \cos(z_2)  \\
0 & 1 & 0 
\end{pmatrix} , \quad
R_1 =
\begin{pmatrix}
1 & 0 & 0 \\
0 & \cos(z_1) &  -\sin(z_1)  \\
0 & \sin(z_1)  & \cos(z_1)  \\
\end{pmatrix} \, ,
\end{equation}
where $S$ is a diagonal matrix of signs and the $z_i$ angles are
complex, hence implying that the $R$ matrix contains 6 real
parameters.

\section{Proof special case: antisymmetric $\boldsymbol{y_1}$ Yukawa matrix}
\label{sec:proof_antisym1}

We consider the special case of an antisymmetric Yukawa matrix, $y_1 =
- y_1^T$, with a general $y_2$ Yukawa. A well-known model with this
feature is the Zee model~\cite{Zee:1980ai}. For simplicity, we focus
on $n_1=n_2=n=3$. We define the invertible matrix
\begin{equation}
H=\Sigma^{1/2}\,V_1\,U^*\,\bar{D}_{\sqrt{m}} \, ,
\end{equation}
and introduce $\bar{H}=H^{-1}\,\widehat{W}$, so that
$H^{-1}=\bar{H}\,\widehat{W}^\dagger$. With these definitions, the
condition $y_1^T+y_1 =0$ is equivalent to
\begin{equation}
y_1^T+y_1 =0 \Leftrightarrow 
(H^{-1}\,W\,T\,C_1)^T + H^{-1}\,W\,T\,C_1=0 \Leftrightarrow 
\begin{pmatrix} C_1^T\,T^T & 0 \end{pmatrix} \, \bar{H}^T 
+ \bar{H} \, \begin{pmatrix} T\,C_1 \\ 0 \end{pmatrix} = 0 \, .
\end{equation}
Since $r = \text{rank}(W) = \text{rank}(y_1)$, the antisymmetry of
$y_1$ implies that $r = 1$ or $r = 2$. We now consider the different
values that $r_m$ and $r$ may take. For each case, we use the form for
$C_1$ and $T$ given in Sec.~\ref{subsec:param}, impose the
antisymmetry condition on $y_1$, and derive expressions for $T$ and
$C_1$ in terms of $\bar{H}$. This leads to several conditions on $T$
and $C_1$ as well as on $\bar{H}$, which we now list. \\

{\bf Case $\boldsymbol{(3,2)_a}$}: In this case $\bar{H}$ must have
the form
\begin{equation}
\bar{H}=\begin{pmatrix} \bar{h}_{11} & \bar{h}_{12} & \bar{h}_{13} \\
0 & \bar{h}_{22} & \bar{h}_{23} \\ 1 & \bar{h}_{32} & \bar{h}_{33} \end{pmatrix} \, ,
\end{equation}
with $\bar{h}_{11} \neq 0 \neq \bar{h}_{32}$, $\bar{h}_{22}<0$ and
$(\bar{h}_{12}-\bar{h}_{32}\bar{h}_{11})^2 +
\bar{h}_{11}^2\bar{h}_{22}^2 + \bar{h}_{22}^2=0$. One also finds
\begin{equation}
T = t
\,\begin{pmatrix} -\bar{h}_{22} & -\bar{h}_{32} \\ 0 &
1 \end{pmatrix} \, ,
\end{equation}
with $t>0$, and $C_1$, given in Eq.~\eqref{eq:C1C232a}, with
$z_1=(\bar{h}_{12}-\bar{h}_{32}\bar{h}_{11})/\bar{h}_{22}$ and
$z_2=\bar{h}_{11}$.  Moreover, these conditions translate into
restrictions on the parameters in $m$ and $M$, since $H\,\bar{H}$ must
be a unitary matrix. \\

{\bf Case $\boldsymbol{(2,2)_a}$}: The matrix $\bar{H}$ can be written
in this case as
\begin{equation}
\bar{H}=\begin{pmatrix} \bar{h}_{11} & \bar{h}_{12} & \bar{h}_{13} \\
0 & \bar{h}_{22} & \bar{h}_{23} \\ \bar{h}_{31} & \bar{h}_{32} & \bar{h}_{33} \end{pmatrix} \, ,
\end{equation}
with $\bar{h}_{31} \neq 0 \neq \bar{h}_{22}$,
$\bar{h}_{31}/\bar{h}_{22}<0$ and
$(-\bar{h}_{11}\bar{h}_{32}+\bar{h}_{12}\bar{h}_{31})^2 +
\bar{h}_{11}^2\bar{h}_{22}^2 = 0$. $T$ is given by
\begin{equation}
T= t\,\begin{pmatrix} -\bar{h}_{22}/\bar{h}_{31} & -\bar{h}_{32}/\bar{h}_{31} \\ 
0 & 1 \end{pmatrix} \, ,
\end{equation}
with $t>0$, and $C_1$ given in Eq.~\eqref{eq:C1C222a}, particularized
with
$z_1=(\bar{h}_{12}\bar{h}_{31}-\bar{h}_{32}\bar{h}_{11})/(\bar{h}_{22}
\bar{h}_{31})$ and $z_2=\bar{h}_{11}/\bar{h}_{31}$.  Again, these
conditions translate into restrictions on $m$ and $M$ since
$H\,\bar{H}$ has to be unitary. \\

{\bf Cases $\boldsymbol{(3,2)_b}$ and $\boldsymbol{(2,2)_b}$}: In
these two cases the form of $C_1$ is common and it does not contain
any parameter, see Eqs.~\eqref{eq:C1C232b} and
\eqref{eq:C1C222b}. Therefore, the resulting expression for $\bar{H}$
is also the same,
\begin{equation}
\bar{H}=\begin{pmatrix} \bar{h}_{11} & \bar{h}_{12} & \bar{h}_{13} \\
0 & \pm i\, \bar{h}_{32} & \bar{h}_{23} \\ 0 & \bar{h}_{32} & \bar{h}_{33} \end{pmatrix} \, ,
\end{equation}
with the conditions $\bar{h}_{11} \neq 0 \neq \bar{h}_{22}$ and
Im$(\mp\bar{h}_{11}/\bar{h}_{22})<0$. In addition, the matrix $T$ is
given by
\begin{equation}
T= t\,\begin{pmatrix} 1 & \pm i\,\bar{h}_{12}/\bar{h}_{22} \\ 
0 & \mp i\,\bar{h}_{11}/\bar{h}_{22}\end{pmatrix} \, ,
\end{equation}
with $t>0$. Finally, the product $H\,\bar{H}$ must be a unitary
matrix, and this again implies restrictions on the entries of the
matrices $m$ and $M$. \\

{\bf Case $\boldsymbol{(2,1)}$}: following the same procedure as in
the previous cases, one concludes in this case that $\bar{H}$ is a
non-invertible matrix, hence finding a contradiction. Therefore, we
discard this possibility in this special case.

\section{Proof special case: $\boldsymbol{\sqrt{2} \,y_1 = \sqrt{2} \,y_2 = y = -y^T}$}
\label{sec:proof_antisym2}

We consider the special case of equal and antisymmetric Yukawa
matrices, $\sqrt{2} \,y_1=\sqrt{2} \,y_2=y=-y^T$. This scenario takes
place in the Zee-Babu~\cite{Cheng:1980qt,Zee:1985id,Babu:1988ki} and
KNT~\cite{Krauss:2002px} models and the 331 model in
\cite{Boucenna:2015zwa}, to mention a few representative
examples. This case necessarily requires $n_1=n_2=3$ and $V_1=V_2
\equiv V$. Furthermore, the antisymmetry of the $y$ Yukawa matrix
implies $r=r_m=2$, and then one of the neutrinos remains massless. For
simplicity, we will restrict our analysis to $n=3$. The condition
$y_1=y_2$ is equivalent to
\begin{equation} \label{eq:y1y2}
y_1=y_2
\Leftrightarrow W \, A = \widehat{W}^* \, \widehat{B} = W^* \, B + \bar{W}^* \, \bar{B} 
\Leftrightarrow \left\{ \begin{array}{l}
\bar{B}=0 \, , \\
W^T W \, A = B  \, . \end{array} \right.
\end{equation}
We define the matrix $R = W \, T$, in the same way as in the type-I
seesaw, see Sec.~\ref{subsec:typeI}. Then Eq.~\eqref{eq:y1y2} is equivalent to
\begin{equation} \label{eq:y1y2b}
y_1=y_2 \Leftrightarrow \left\{ \begin{array}{l}
\bar{B}=0 \, , \\
R^T R \, C_1 = C_1 \, C_2 + K\, C_1 \, . \end{array} \right.
\end{equation}
Since $r = r_m = 2$, in principle one has two possible scenarios: case
$(2,2)_a$ and case $(2,2)_b$. The latter is not compatible with
Eq.~\eqref{eq:y1y2b}, since the components $(1,2)$ and $(2,2)$ of $R^T
R \, C_1 = C_1 \, C_2 + K\, C_1$ leads to $-1=1$. In contrast, case
$(2,2)_a$ is perfectly compatible with Eq.~\eqref{eq:y1y2b}. Now $R^T
R \, C_1 = C_1 \, C_2 + K\, C_1$ leads to $K=0$ and $z_1=z_2=0$ in the
expression of $C_1$ given in Eq.~\eqref{eq:C1C222a}. One also finds
that $R$ is a $3\times 2$ matrix such that $R^T R = \id_2$. Therefore,
we find a modified Casas-Ibarra parametrization, with
\begin{equation}
y = \sqrt{2} \, y_1 = \sqrt{2} \, y_2 = i \, V^\dagger \, \Sigma^{-1/2} \, R \, C_1 \, \bar{D}_{\sqrt{m}} \, U^\dagger \, ,
\end{equation}
and $R$ a $3\times 2$ Casas-Ibarra matrix. However, we still must impose
the antisymmetry condition on $y$. As we will see, this will imply
non-trivial restrictions on the $R$ and $\Sigma$ matrices, which can
no longer be general. First, we define
\begin{align}
F &= \Sigma^{-1/2} \, V^*\, U \, \bar{D}_{\sqrt{m}}^{-1} \, C_1^T \, , \\
\bar{F} &= \Sigma^{1/2} \, V \, U^* \, \bar{D}_{\sqrt{m}} \, C_1^T \, ,
\end{align}
two $3\times 2$ matrices of rank 2 which satisfy $\bar{F}^T F =
\id_2$. With these definitions, one finds
\begin{equation}
y^T+y=0 \Leftrightarrow R \, \bar{F}^T + \bar{F} \, R^T=0 \Rightarrow 
R^T \, F + F^T \, R = 0 \, .
\end{equation}
Next, we introduce the vector $f_3$, such that $f_3^T F =
0_{1\times 2}$ and $f_3^T f_3 = 1$. Therefore, the columns of the
matrix $\begin{pmatrix} F & f_3 \end{pmatrix}$ are a basis of
$\mathbb{C}^3$ and $f_3$ is defined up to a sign. Since
\begin{equation}
\begin{pmatrix}\bar{F} & f_3 \end{pmatrix}^T \begin{pmatrix} F & f_3 \end{pmatrix} 
= \begin{pmatrix} \id_2 & \bar{F}^T f_3 \\ 0 & 1 \end{pmatrix}
\end{equation}
is an invertible matrix, $\begin{pmatrix}\bar{F} & f_3 \end{pmatrix}$
also forms a basis of $\mathbb{C}^3$. We also define the vector $r_3$
such that $\widehat{R}=(R \;\; r_3)$ is an orthogonal matrix, with
$R^T r_3=0_{2\times 1}$ and $r_3^T r_3 =1$. We now write $F$ in terms
of the basis $\widehat{R}$, $F=\widehat{R}\,\widehat{G}$, with
\begin{equation}
\widehat{G}=\begin{pmatrix} G \\ g_3^T \end{pmatrix}
\end{equation}
a $3\times 2$ matrix, $G$ a $2\times 2$ matrix and $g_3$ a
2-components vector. With these definitions, the antisymmetry of the
$y$ Yukawa matrix translates into
\begin{equation}
y^T+y = 0 \Leftrightarrow G^T + G = 0 \, ,
\end{equation}
and $G$ is an antisymmetric matrix. Since $F = \widehat{R} \,
\widehat{G} = R \, G + r_3 \, g_3^T$ has rank 2, and $G$ is
antisymmetric, $G\neq 0$ and therefore $G$ is invertible. This allows
us to write 
\begin{equation}
G = \begin{pmatrix}0 & G_{12} \\ - G_{12} & 0 \end{pmatrix}
\end{equation}
and $R = (F - r_3 \, g_3^T) \, G^{-1}$. The condition $R^T
r_3=0$ is equivalent to $g_3=F^T r_3$, and then $R$ has the form $R =
(F - r_3 \, r_3^T\, F) \, G^{-1}$. We now write $r_3$ in terms of the
basis $\begin{pmatrix}\bar{F} & f_3 \end{pmatrix}$. For this purpose,
we define
\begin{equation} \label{eq:defr3}
r_3 = \begin{pmatrix}\bar{F} & f_3 \end{pmatrix} \begin{pmatrix} \alpha \\ \alpha_3 \end{pmatrix} \, ,
\end{equation}
with
\begin{equation} \label{eq:defalpha}
\alpha = \begin{pmatrix} \alpha_1 \\ \alpha_2 \end{pmatrix} \, ,
\end{equation}
and $\alpha_i\in\mathbb{C}$. We note that the freedom in the global
sign of $f_3$ can be absorbed in $\alpha_3$. From the definition in
Eq.~\eqref{eq:defr3}, it follows that $r_3^T F = \alpha^T$ and the
$R$ matrix can be rewritten as
\begin{equation} \label{eq:Rantisym2}
R = (F - r_3 \, \alpha^T) \, G^{-1} \, .
\end{equation}
This form for the $R$ matrix is a necessary but not sufficient
condition to guarantee the antisymmetry of $y$, which has not been
fully established yet. Three conditions must
be satisfied:

\begin{enumerate}[label=(\roman*)]
\item $y^T+y = 0$,
\item $R^T R = \id_2$,
\item $r_3^T r_3=1$.
\end{enumerate}

In the following we build on these conditions and use them to compute
explicitly $G$ and the $\alpha_i$ parameters, with $i=1,2,3$. The
combination of the $R$ matrix in Eq.~\eqref{eq:Rantisym2} and the
resulting expressions will constitute the most general solution to $m
= y^T M y$ with an antisymmetric $y$ matrix. First, we note that
condition (ii) is equivalent to $F^T F + G^2 = F^T \, r_3 \, r_3^T \,
F = \alpha \, \alpha^T$. Now, the antisymmetry condition (i) can be
used together with Eq.~\eqref{eq:Rantisym2} to derive
\begin{equation}
y^T+y=0 \Leftrightarrow R \, \bar{F}^T + \bar{F} \, R^T=0 \Leftrightarrow 
(F-r_3\,r_3^T\,F) \, G^{-1} \bar{F}^T - \bar{F}\,G^{-1} (F^T-F^T\,r_3\,r_3^T)=0 \, .
\end{equation}
Multiplying on the left by $\begin{pmatrix} F^T
  \\ f_3^T \end{pmatrix}$ and on the right by $\begin{pmatrix} F &
  f_3 \end{pmatrix}$, invertible matrices in both cases, and taking
into account $r_3^T F = \alpha^T$ and condition (ii), we get after some
simplifications
\begin{equation}
y^T+y = 0 \Leftrightarrow \begin{pmatrix} 0 & B_1 \\ B_1^T & B_2 \end{pmatrix} = 0 \, ,
\label{eq:antisim}
\end{equation}
where 
\begin{align}
B_1 &= G^{-1} \, F^T \, F \, \bar{F}^T \, f_3 + G^{-1} \alpha_3 \, \alpha \, , \\
B_2 &= -(f_3^T \, \bar{F} \, F^T \, F + \alpha_3 \alpha^T) G^{-1} \bar{F}^T f_3 
+ f_3^T \, \bar{F} \, G^{-1} (F^T \, F \, \bar{F}^T \, f_3 + \alpha_3 \, \alpha) \, .
\end{align}
It is easy to see that Eq.~\eqref{eq:antisim} is equivalent to $F^T \,
F \, \bar{F}^T \, f_3 + \alpha_3 \, \alpha = 0$. Therefore,
in summary, conditions (i)-(iii) are equivalent to:

\begin{enumerate}[label=(\roman*)]
\item $F^T \, F \, \bar{F}^T \, f_3 + \alpha_3 \, \alpha = 0$,
\item $F^T \, F = G^2 + \alpha \alpha^T$,
\item $r_3^T r_3=1$.
\end{enumerate}

These three conditions are better suited to find $G$ (or,
equivalently, $G_{12}$) and the $\alpha_i$ parameters. We define
$L=F^T F$, $\bar{L}=\bar{F}^T\bar{F}$ and $\bar{\bar{L}}=\bar{F}^T
f_3$. We distinguish two possibilities.

\begin{itemize}
\item $L_{12}\neq 0$
\end{itemize}

If $L_{12}\neq 0$, it is straightforward to use conditions (i)-(iii)
to explicitly compute $G_{12}$ and the $\alpha_i$ parameters. We find
\begin{align}
\alpha_1 &= \sqrt{\frac{L_{12} (L\,\bar{\bar{L}})_{11}}{(L\,\bar{\bar{L}})_{21}}} \, , \\
\alpha_2 &= \frac{L_{12}}{\alpha_1} \, , \\
\alpha_3 &= - \frac{(L\,\bar{\bar{L}})_{11}}{\alpha_1} \, , \\
G_{12} &= \epsilon \, \sqrt{L_{22}-\alpha_2^2} \, ,
\end{align}
with $\epsilon=\pm1$. In addition, one finds two non-trivial
restrictions on the parameters of the model, given by
\begin{align}
\alpha_1^4 \, \bar{L}_{11} + \alpha_1^2 \, [-1 + 2 L_{12} \bar{L}_{12} - 2 \bar{\bar{L}}_{11} (L\,\bar{\bar{L}})_{11}] + [L_{12}^2 \bar{L}_{22} - 2 L_{12} \bar{\bar{L}}_{21} (L\,\bar{\bar{L}})_{11} + (L\,\bar{\bar{L}})_{11}^2] &= 0 \, , \\
\alpha_1^4 + \alpha_1^2 (L_{22}-L_{11})-L_{12}^2 &= 0 \, .
\end{align}

\begin{itemize}
\item $L_{12} = 0$
\end{itemize}

If $L_{12}=0$, three solutions exist:

\begin{itemize}[label=$\ast$]
\item Solution 1:
\end{itemize}

\begin{align}
\alpha_1 &=0 \, , \\
\alpha_2 &= \sqrt{L_{22}-L_{11}}\neq 0 \, , \\
\alpha_3 &= -\frac{L_{22} \, \bar{\bar{L}}_{21}}{\alpha_2} \, , \\
G_{12} &= \epsilon \, \sqrt{L_{11}} \, ,
\end{align}
with the conditions $L_{11} \neq 0$, $\bar{\bar{L}}_{11}=0$ and
$(L_{22}-L_{11}) \, \bar{L}_{22} - 2 L_{22} \, \bar{\bar{L}}_{21} =
1$.

\begin{itemize}[label=$\ast$]
\item Solution 2:
\end{itemize}

\begin{align}
\alpha_2 &= 0 \, , \\
\alpha_1 &= \sqrt{L_{11}-L_{22}}\neq 0 \, , \\
\alpha_3 &= - \frac{L_{11} \, \bar{\bar{L}}_{11}}{\alpha_1} \, , \\
G_{12} &= \epsilon \, \sqrt{L_{22}} \, ,
\end{align}
with the conditions $L_{22} \neq 0$, $\bar{\bar{L}}_{21}=0$ and
$\displaystyle (L_{11}-L_{22}) \, \bar{L}_{11} - 2 L_{11} \,
\bar{\bar{L}}_{11} + \frac{L_{11}^2 \,
  \bar{\bar{L}}_{11}^2}{L_{11}-L_{22}} = 1$.

\begin{itemize}[label=$\ast$]
\item Solution 3:
\end{itemize}

\begin{align}
\alpha_1 &= \alpha_2 = 0 \, , \\
\alpha_3 &= 1 \, , \\
G_{12} &= \epsilon \, \sqrt{L_{11}} = \epsilon \, \sqrt{L_{22}} \, ,
\end{align}
with the conditions $L_{22}-L_{11} = \bar{\bar{L}}_{11} =
\bar{\bar{L}}_{21} = 0$.

\section{Yukawa parametrization, loop corrections and fine-tuning}
\label{sec:finetuning}

In this Appendix we discuss how fine-tunings in the parametrization of
the Yukawa matrices might be spoiled by higher-order loop corrections.
We will then demonstrate how one can easily take these contributions
into account in Eq.~\eqref{eq:master}, such that neutrino masses (and
angles) remain correctly fitted, even in such particularly sensitive
parts of the parameter space. In this discussion, we will use the
simplest type-I seesaw with three right-handed neutrinos as an
example. For other models one can use a similar, albeit in some cases
more involved procedure.

In the main text we have treated the parameters entering the various
matrices $W$, $A$, $B$ and so on as completely free parameters.
Nevertheless, physically there are restrictions on these parameters from
the requirement that the Yukawa couplings do not enter the
non-perturbative regime.  It is, of course, easy to check that all
$|y_{ij}|$ are smaller than some critical value $y_{\rm cr}$, say for
example $y_{\rm cr} \le \sqrt{4\pi}$, for any given choice of the other
free parameters.

However, even for Yukawa couplings $|y_{ij}|\ll 1$, the
tree-level formulas may fail in some regions of parameter space. 
Consider the total neutrino mass matrix $m$, written as
\begin{equation}\label{eq:TrPlusLp}
m = m^{\rm Tree} + \delta m^{\rm 1-loop} + \cdots
\end{equation}
Here, the dots stand for higher order corrections, while the
superscripts ${\rm Tree}$ and ${\rm 1-loop}$ indicate the tree-level
and 1-loop contributions to $m$. It is natural to asume that $\delta
m^{\rm 1-loop}/m^{\rm Tree} \ll 1$, which for seesaw type-I is true in
most parts of parameter space, but not in a particular region, on
which we will from now on concentrate.~\footnote{Recall that precision
  global fits \cite{deSalas:2017kay} now give error bars for $\Delta
  m^2_{ij}$ of a few percent only. Thus, even small loop terms might
  induce numerically important shifts in the final result.}

As explained in Sec.~\ref{subsec:typeI}, in the type-I seesaw with
three right-handed neutrinos the neutrino mass matrix is given {\em at
  tree-level} by $m = - \frac{v^2}{2} \, y^T \, {M_N}^{-1} \, y$, an
expression that can be obtained with the master formula by taking
$f=-1$, $n_1 = n_2 = 3$, $y_1 = y_2 = y/\sqrt{2}$ and $M =
\frac{v^2}{2} M_N^{-1}$. In this minimal type-I seesaw, one can always
go to a basis where the mass matrix of the right-handed neutrinos is
diagonal, $M_N \to {\widehat M}_N$, with eigenvalues $m_{N_i}$, which are
free parameters. In this basis, the master parametrization reduces to
the well-known Casas-Ibarra parametrization in Eq.~\eqref{eq:CI},
which introduces the $3 \times 3$ orthogonal matrix $R$, parametrized
in Appendix~\ref{sec:mat} in terms of 3 complex angles, see
Eqs.~\eqref{eq:defR} and \eqref{eq:defR2}.

One-loop corrections to the seesaw formula have been calculated
several times in the
literature~\cite{Grimus:2002nk,AristizabalSierra:2011mn}. They can be
written as
\begin{equation}\label{eq:lp}
 \delta m^{\rm 1-loop} = -\frac{v^2}{2} \, y^T \, M_R^{-1} \, {\widehat \Delta}^{\rm Loop} \, y,
\end{equation}
where~\cite{AristizabalSierra:2011mn}
\begin{equation}\label{eq:lp2}
  {\widehat \Delta}^{\rm Loop} =\frac{g^2}{64 \pi^2 m_W^2}
  \left[m_h^2 \, \ln \left( \frac{\widehat M_N^2}{m_h^2} \right)
  + 3 \, m_Z^2 \, \ln \left( \frac{\widehat M_N^2}{m_h^2} \right) \right] \, .
\end{equation}
Note that ${\widehat \Delta}^{\rm Loop}$ is dimensionless and typically of
order per-mille to percent for right-handed neutrino masses of order
${\cal O}(0.1-1)$ TeV. We stress that Eq.~\eqref{eq:lp} has again the
form of the master formula.

Let us parametrize the complex angles in $R$ as~\cite{Anamiati:2016uxp}
\begin{equation}\label{eq:defzi}
z_i = \kappa_i \cdot e^{2i\pi\, \alpha_i} ,
\end{equation}
where $\alpha_i$ are real numbers $\in [0,1]$, and $\kappa \in
[0,\kappa_{\rm max}]$. One can consider the upper limit $\kappa_{\rm
  max}$ as a measure of how much fine-tuning is allowed in the
Yukawas. Maximal fine-tuning (as function of $\kappa_i$) occurs for 
$\alpha_i=1/4$.  

\begin{figure}
 \centering
 \includegraphics[width=0.41\textwidth]{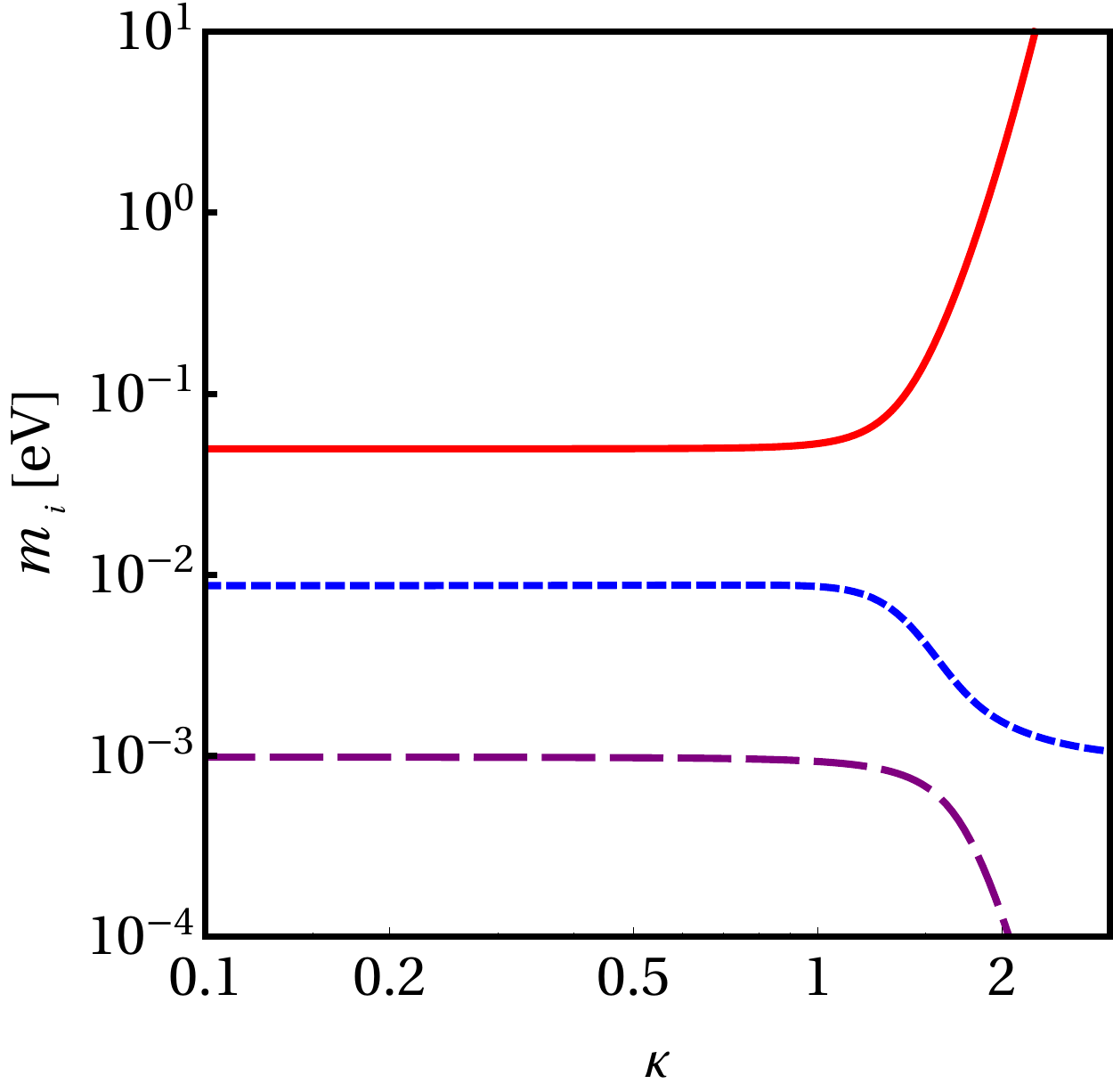}
\hskip5mm
 \includegraphics[width=0.42\textwidth]{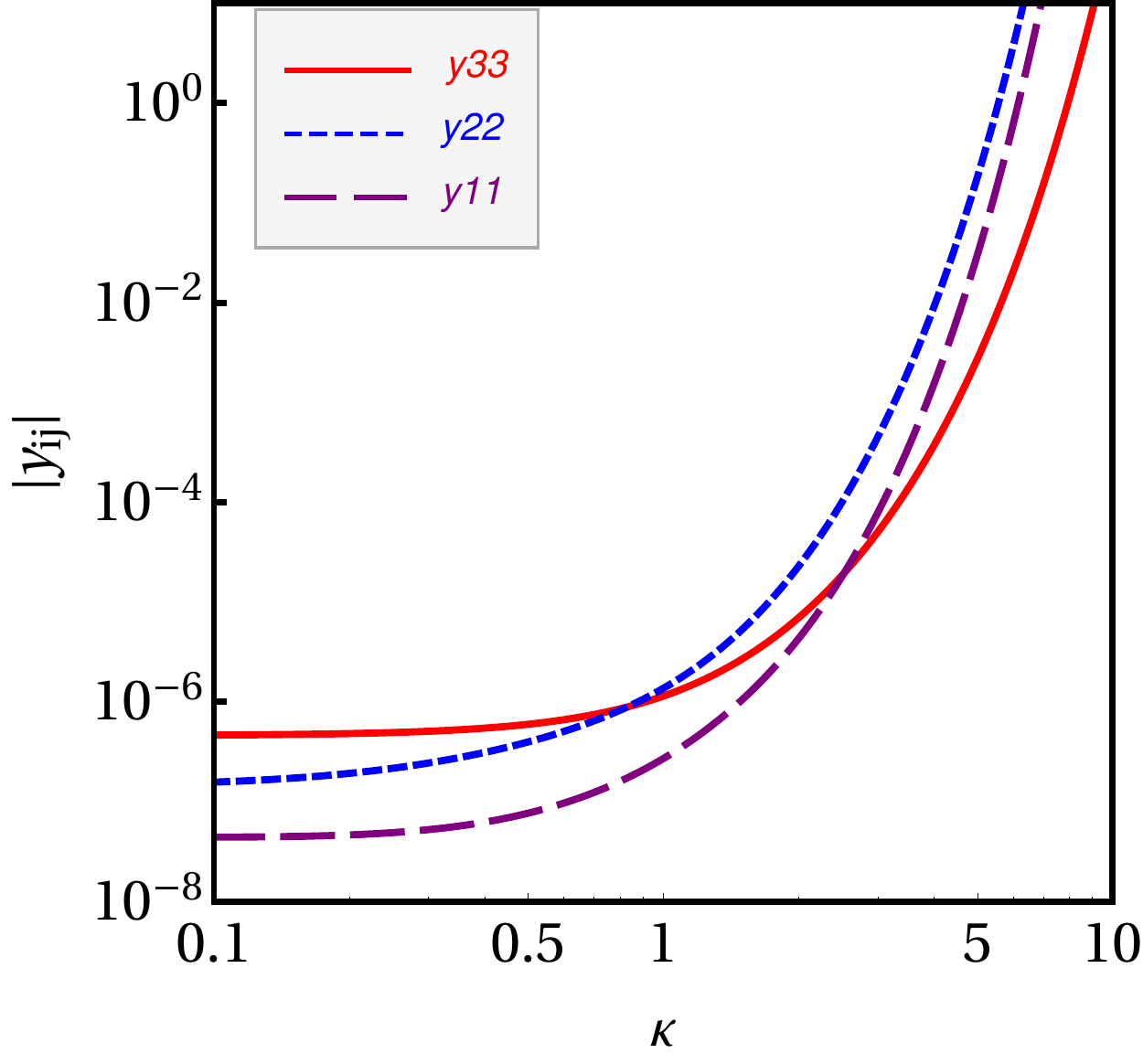}
 \caption{Example neutrino masses versus $\kappa_i=\kappa$, see
   Eq.~\eqref{eq:defzi}, for best-fit oscillation data and one
   particular choice of $m_{1}=10^{-3}$ eV, $m_{N_1}=100$ GeV,
   $m_{N_2}=200$ GeV, $m_{N_3}=300$ GeV, and all phases
   $\alpha_i=1/4$. To the left: Eigenvalues calculated from the
   tree-level expression for the Yukawa couplings, see the
   Casas-Ibarra parametrization in Eq.~\eqref{eq:CI}. To the right:
   Absolute values of the yukawa couplings $y_{11}$, $y_{22}$,
   $y_{33}$ for the same fit.
\label{fig:mnuVkap}}
\end{figure}

On the left-hand side of Fig.~\ref{fig:mnuVkap} we show examples of
the light neutrino masses, calculated from Yukawas as given by the
Casas-Ibarra parametrization in Eq.~\eqref{eq:CI}, for some particular
but random choice of inputs, as a function of $\kappa$, assuming
$\kappa_i=\kappa$. Here, the neutrino mass matrix includes the
loop corrections, while the Casas-Ibarra parametrization is at
tree-level. For $\kappa \le 1$, the neutrino masses are
constant, demonstrating that the fit procedure is stable and the
output neutrino masses equal the input values. However, for $\kappa >
1$, neutrino masses can deviate by orders of magnitude from their
desired values.

One can understand this behavior with the help of the right-hand side
of Fig.~\ref{fig:mnuVkap}. For small values of $\kappa$ the Yukawa
couplings do change, but remain of the same order of
magnitude. Typical values are of the naive order of $y_{ii} \propto
\sqrt{m_{i}/m_{N_i}}$. Increasing $\kappa$ beyond 1 leads to Yukawas
larger than this naive estimate, which indicates that in the neutrino
mass matrix small neutrino masses are generated {\em by a cancellation
  among different terms}. These cancellations are unstable against
radiative corrections, which explains the behavior of the output
neutrino masses for large $\kappa$. We stress that this unwanted
behavior occurs already for Yukawas much smaller than one.

Given that the structure of Eq.~\eqref{eq:lp} is necessarily again of the
same form as the master formula, however, it is straight-forward to
correct Eq.~\eqref{eq:CI}, in order to take the 1-loop
contributions into account. One simply replaces the eigenvalues in
${\widehat M}_R$ in Eq.~\eqref{eq:CI} by
\begin{equation}\label{eq:modmR}
{\widehat M}_R^{-1} \to {\widehat M}_R^{-1} \, (\id + {\widehat \Delta}^{\rm Loop})
\end{equation}
This (small) shift corrects the Yukawa couplings in the right 
way, such that the change of output neutrino masses for 
$\kappa \gsim 1$ disappears. Note, however, that for 
$\kappa$ larger than $\kappa \sim 5$, Yukawas enter the 
non-perturbative regime and the calculation will fail in any 
case.

\section{Hybrid scenarios}
\label{sec:hybrid}

A relatively natural question one can consider is whether it is
possible or not to use our master parametrization in a model with
several contributions to the neutrino mass matrix. More precisely, let
us consider a model leading to a neutrino mass matrix of the form
\begin{equation} \label{eq:hybrid}
m = \sum_i^N \bar m_i = \bar m_1 + \bar m_2 + \dots + \bar m_N \, ,
\end{equation}
where each of the contributions to the neutrino mass matrix is given
by $\bar m_i$.~\footnote{A familiar example of such situation is given by
  supersymmetry with bilinear R-parity violation, a scenario in which
  $\bar m_1$ would correspond to the tree-level contribution, while $\bar m_2$
  would denote the 1-loop contribution. More exotic hybrid examples,
  with completely independent contributions to the total neutrino mass
  matrix, can also be found.} Of course, the strategy is to bring the
sum $\sum_i \bar m_i$ into the form required by our master formula, since
that would make the master parametrization directly applicable. This
can be done in general, as we proceed to illustrate now in a scenario
with two contributions to the neutrino mass matrix, $\bar m_1$ and $\bar m_2$,
each given by a Yukawa matrix, $Y_1$ and $Y_2$. In this case,
Eq. \eqref{eq:hybrid} reduces to
\begin{equation} \label{eq:hybrid2}
m = \bar m_1 + \bar m_2 = Y_1^T \, \mathcal{M}_{11} \, Y_1 + Y_1^T \, \mathcal{M}_{12} \, Y_2 + Y_2^T \, \mathcal{M}_{12}^T \, Y_1 + Y_2^T \mathcal{M}_{22} Y_2 \, .
\end{equation}
In case the two contributions to the total neutrino mass matrix are
completely independent, $\mathcal{M}_{12} = 0$ and $m$ is just given
by $Y_1^T \, \mathcal{M}_{11} \, Y_1 + Y_2^T \mathcal{M}_{22}
Y_2$. However, we consider the possibility of a crossed term, given by
$\mathcal{M}_{12} \neq 0$. Eq. \eqref{eq:hybrid2} can be rewritten as
\begin{equation} \label{eq:hybrid3}
m = y^T \, M \, y  \, ,
\end{equation}
with
\begin{equation}
y = \left( \begin{array}{c}
Y_1 \\
Y_2 \end{array} \right) \quad \text{and} \quad M = \left( \begin{array}{cc}
\mathcal{M}_{11} & \mathcal{M}_{12} \\
\mathcal{M}_{12}^T & \mathcal{M}_{22} \end{array} \right) \, ,
\end{equation}
and this is formally equivalent to the master formula in
Eq. \eqref{eq:master}, which in turn implies that the master
parametrization can be directly applied.  This procedure can be easily
generalized to hybrid scenarios with more than two contributions
(independent or not) to the total neutrino mass matrix. We mention,
however, that in case $Y_1$ and/or $Y_2$ have to fullfil some
particular constraints, application of the master parametrization may
not be straighforward anymore, as discussed in Section~\ref{sec:extra}.

\providecommand{\href}[2]{#2}\begingroup\raggedright\endgroup

\end{document}